\documentclass[12pt]{article}
\usepackage{amssymb}
\usepackage{amsmath}
\usepackage{amstext}
\usepackage{graphicx,epsfig}
\usepackage{epsfig}
\usepackage{color}
\usepackage{parskip}
\usepackage{cite}
\usepackage{mathtools}
\usepackage[utf8]{inputenc}

\allowdisplaybreaks

\newcommand{\Comment}[1]{{}}
\definecolor{MyDarkBlue}{rgb}{0.15,0.15,0.45}
\usepackage[linktocpage=true]{hyperref}
\hypersetup{
colorlinks=true,
citecolor=MyDarkBlue,
linkcolor=MyDarkBlue,
urlcolor=MyDarkBlue,
pdfauthor={Kurt Hinterbichler},
pdftitle={},
pdfsubject={hep-th}
}

\newcommand{\be}{\begin{equation}}
\newcommand{\ee}{\end{equation}}
\newcommand{\bal}{\begin{align}}
\newcommand{\eal}{\end{align}}
\newcommand{\bals}{\begin{align*}}
\newcommand{\eals}{\end{align*}}
\newcommand{\bea}{\begin{eqnarray}}
\newcommand{\eea}{\end{eqnarray}}
\newcommand{\beas}{\begin{eqnarray*}}
\newcommand{\eeas}{\end{eqnarray*}}

\def\({\left(}
\def\){\right)}

\DeclarePairedDelimiter\floor{\lfloor}{\rfloor}

\numberwithin{equation}{section}

\begin{document}
\begin{center}
{\LARGE {Bounds on Amplitudes in Effective Theories with Massive Spinning Particles \\ \vspace{.2cm} }}
\end{center}
\vspace{2truecm}
\thispagestyle{empty} \centerline{
{\large {James Bonifacio,}}$^{}$\footnote{E-mail: \Comment{\href{mailto:jjb239@case.edu}}{\tt james.bonifacio@case.edu}}
{\large { Kurt Hinterbichler$^{}$}}\footnote{E-mail: \Comment{\href{mailto:kurt.hinterbichler@case.edu}}{\tt kurt.hinterbichler@case.edu}}
}

\vspace{1cm}

\centerline{\it ${}^{\rm }$CERCA, Department of Physics, Case Western Reserve University, }
\centerline{\it 10900 Euclid Ave, Cleveland, OH 44106, USA}

\begin{abstract}

We consider a procedure for directly constructing general tree-level four-particle scattering amplitudes of massive spinning particles that are consistent with the usual requirements of Lorentz invariance, unitarity, crossing symmetry, and locality.  There are infinitely many such amplitudes, but we can isolate interesting theories by bounding the high-energy growth of the tree amplitudes within the effective field theory. This allows us to set model-independent lower bounds on the growth of tree-level amplitudes in any effective field theory with a given particle content and any interaction terms with an arbitrary but finite number of derivatives.  In certain common cases this corresponds to finding the highest possible strong coupling scale. 
When applied to spin 2, we show that the only amplitudes that saturate this bound are generated by the known ghost-free theories of a massive spin-2 particle, namely dRGT massive gravity and the pseudolinear theory.
We also make a conjecture for the allowed growth of tree amplitudes in a theory with a single massive particle of any integer spin.

\end{abstract}

\newpage

\setcounter{tocdepth}{3}
\tableofcontents

\section{Introduction}
\parskip=5pt
\normalsize

Interacting massive higher-spin particles exist---both theoretically and in nature---as ingredients in consistent fundamental theories.  In nature, the heavy hadronic resonances in quantum chromodynamics (QCD) carry high spin and and interact strongly with each other.  In string theory, the higher vibrational modes of the string are massive higher-spin particles, and can be weakly interacting if the string coupling is small.  

In all known cases, however, a theory containing a finite number of interacting massive high spin fields is always an effective theory; there is some intrinsic strong coupling scale beyond which tree amplitudes violate perturbative unitarity bounds, meaning that strong coupling effects or new particles must come in at a scale not greater than the scale of unitarity violation.  In the case of the higher spin hadrons they become strongly coupled at the QCD scale, the intrinsic size of the hadrons, where they fail to be pointlike and require strong dynamics to complete the description.  But this is also the scale that sets the mass of the hadrons, so there is no regime in which they are well described by a point-particle effective field theory (EFT) with a strong coupling scale parametrically larger than the mass.  In the case of string theory, the higher-spin states can be weakly interacting, but the description requires an infinite tower of higher spins, and the masses within the tower are not parametrically separated from each other, so that again there is no point-like description of a finite number of massive fields with a cutoff parametrically higher than the mass itself.

The fact that these particular ultraviolet (UV) complete examples fail to have clean EFT descriptions which include a finite number of massive states does not imply that any possible UV theory will fail in the same way.  There are strong no-go results against massless high spins in flat space (as reviewed in, e.g. \cite{Bekaert:2010hw,Porrati:2012rd}), but these do not apply to massive particles, and comparatively little is known about constraints on possible massive high spin interactions.  

Purely at the level of EFTs, it is not difficult to write down examples of massive interacting higher-spin particles with strong coupling scales parametrically higher than the cutoff.  
In the case of spin 1, a canonical example is the self-interacting theory of massive spin-1 particles, like the $W$ and $Z$ bosons of the standard model.  In the Abelian case, the tree-level amplitude for four-point scattering of longitudinal modes grows with energy like $E^4$. This growth comes with the scale $\Lambda\sim m/g$, where $g$ is a dimensionless coupling constant and $m$ the mass of the spin-1 particle.  If $g$ is small, then the strong coupling scale is parametrically larger than the mass and we have a consistent EFT description of the massive spin-1 particle up to at most the scale $\Lambda$.  In the Abelian Higgs model \cite{Englert:1964et,Higgs:1964pj}, a weakly coupled Higgs boson comes in at a mass scale $\mu\sim {\lambda^{1/2}}\Lambda$, where $\lambda$ is the coupling constant characteristic of the weakly coupled UV completion (the Higgs quartic coupling).  To the extent that $\lambda$ is small, the Higgs boson comes in before the strong coupling scale and prevents perturbative unitarity violation.   

One might wonder if it is possible to tune the effective theory in some way, by adding the right interactions, but without adding new degrees of freedom, to reduce the growth of the amplitude.  As we will see, this is not possible: the $E^4$ behavior of the amplitude is both the generic and the best possible for an EFT of a single massive spin-1 particle. 

Moving to spin 2, there are some new subtleties and possibilities for the effective theory, and it is indeed possible to improve the growth of the amplitude somewhat.
As in the spin-1 case, it is easy to write effective theories with strong coupling scales parametrically higher than the mass.  For example, simply adding generic zero-derivative mass terms to the Einstein-Hilbert action results in an EFT whose tree-level four-particle amplitude grows with energy as $E^{10}$.  With standard choices for scalings of the fields and derivatives, this implies a strong coupling scale of $\Lambda_5 \equiv (M_Pm^4)^{1/5}$, where $M_P$ is the Planck mass and $m$ the spin-2 mass \cite{ArkaniHamed:2002sp}.
However, it was shown in Refs.~\cite{ArkaniHamed:2002sp,Creminelli:2005qk} that there exist particular choices of mass terms that improve the growth of the amplitude to $E^6$, with a concomitant raised strong coupling scale $\Lambda_3 \equiv (M_Pm^2)^{1/3}$.  This choice leads to the ghost-free massive gravity theories of de Rham, Gabadadze and Tolley (dRGT) \cite{deRham:2010ik,deRham:2010kj}.  

The improved behavior of the dRGT amplitudes occurs because of a cancellation between the high-energy parts of the exchange diagrams and contact diagrams.  This has no analog in the case of a single spin-1 particle. It was shown in Ref.\cite{Schwartz:2003vj} that if a theory is constructed by adding zero-derivative terms to the Einstein-Hilbert term, with no additional derivative interactions, then the $E^6$ behavior is in fact the best possible for the amplitude, i.e. the dRGT choice produces amplitudes with the slowest growth among theories whose derivative interations are fixed to the Einstein-Hilbert form.

However, this $E^6$ behavior has not yet been shown to be a truly model-independent lower bound, because there is still the possibility that other derivative interactions, beyond those of the Einstein-Hilbert form, could be used to further slow the growth of the amplitude.\footnote{It was shown in Ref.~\cite{deRham:2013tfa} that there are no additional parity-conserving ghost-free terms that can be added to the dRGT interactions, so the Einstein-Hilbert kinetic term is known to be unique from the point of view of ghost freedom. Our analysis with amplitudes is insensitive to ghosts.}
Here we will rule out this possibility and show that $E^6$ is a true lower bound on the growth of amplitudes for EFTs of a single interacting massive spin-2 field with a finite number of interactions.  We will show that regardless of the choice of interaction terms with any finite number of derivatives, ghostly or not, parity violating or not, there is no way to further reduce the growth of the four-point amplitude without it vanishing completely.  Furthermore, we will see that the only way to saturate the lower bound of $E^6$ is through dRGT massive gravity or the pseudolinear interactions of Refs.~\cite{Folkerts:2011ev, Hinterbichler:2013eza}. This conclusion means that dRGT is unique up to quartic order in the fields, insofar as it is the only effective theory of a single massive spin-2 particle with an Einstein-Hilbert limit that achieves the lower bound on the growth.  Any claims of nonuniqueness or additional interactions \cite{Kimura:2013ika, Wittner:2018jrq} must either not describe a pure massive spin-2 particle in flat spacetime, have amplitudes that grow faster with energy, or differ only at the quintic order and above. 

The caveat of a finite number of derivatives is an important one. Given a UV complete theory with only massive particles, the growth of the amplitude at high energies is bounded above by the Froissart-Martin bound \cite{Froissart:1961ux, Martin:1962rt}.  Integrating out a particle with a large mass will produce an effective theory for the lighter particles.  The effective theory will be organized in part by a derivative expansion, with higher powers of derivatives suppressed by powers of the mass of the particle that was integrated out.  As far as the scattering of light particles is concerned, keeping this entire tower of derivative interactions is equivalent to keeping the intermediate massive state, and so the amplitude as calculated with the entire tower does not grow too much with energy.  As we go up in order in the derivative expansion the amplitude behaves worse and worse at high energies, but once the entire tower is resummed the amplitude becomes consistent with the Froissart-Martin bound.  Thus by allowing for an infinite number of derivatives we would expect to be able to get an amplitude consistent with this bound by arranging for a resummation into a function of energy that falls off at infinity, which would be saying something about the UV completion. We are instead concerned with cancellations that may occur with only a finite number of derivatives, i.e. in a truncated EFT.  Thus the high-energy growth is always polynomial and we are saying nothing about the existence or properties of a putative UV completion (except that the theory cannot be completed by adding a finite number of terms to the effective theory).

Our results are in the same spirit as Refs.~\cite{Porrati:2008an,Porrati:2008ha}, where model-independent strong coupling scales are found for theories of massive higher-spin particles coupled to electromagnetism.  In these works, the model-independent cutoff $\Lambda\sim  {m e^{-1/(2l-1)}}$ is found for a spin-$l$ particle with mass $m$ interacting with electromagnetism with strength $e$, assuming the presence of the minimal coupling interaction. 
This analysis was done at the level of the Lagrangian, by looking for the scale suppressing various possible nonremovable interaction terms. 
Our approach is instead to study directly the on-shell scattering amplitudes by writing down the most general amplitude consistent with the requirements of Lorentz invariance, unitarity, crossing symmetry, and locality.  This method of constructing amplitudes directly, without recourse to a Lagrangian, can be thought of as a kind of bootstrap procedure, and it may prove useful for other problems beyond the specific application of finding a lower bound on the growth that we focus on here. 
 
\subsection{Summary and outline\label{methodsec}}

We now briefly summarize the problem we want to solve and outline our approach.
Given some particles in an EFT with operators containing derivatives up to some arbitrary but finite order, the general problem is to find the minimum nonzero value of
\be \label{eq:ndef}
n \equiv \lim_{E \rightarrow \infty} \frac{d \ln \mathcal{A}(E)_{\rm EFT}}{d \ln E},  
\ee
where $\mathcal{A}(E)_{\rm EFT}$ is the 2-to-2 tree amplitude calculated in the EFT as a function of the center-of-mass energy $E$.  When there are multiple degrees of freedom, we minimize $n$ for any choice of external states.
Of course, in an EFT the tree-level amplitudes containing only the low-energy degrees of freedom are not valid up to arbitrarily high energies. The exponent $n$ measures the high-energy behavior of the amplitude inferred from the EFT, whereas the actual physical amplitude above the cutoff is described by some unknown UV completion. Minimizing $n$ in Eq.~\eqref{eq:ndef} thus corresponds to improving the inferred high-energy behavior of the amplitude as much as possible in the EFT with a finite number of operators.\footnote{If the amplitude is such that higher powers of energy are suppressed by smaller scales, then minimizing $n$ also corresponds to maximizing the strong coupling scale of the effective theory. This is the case, for example, if we have a non-zero cubic vertex for a massive particle schematically of the form $\partial^2 h^3/M_p$, where $m$ is the mass of the particle and $M_p$ is some mass scale such that $m \ll M_p$. By dimensional analysis, it follows that for $E \gg m$ the four-point amplitude can be expanded as a sum of terms of the form $E^{n}/M_p^2 m^{n-2}$.
Terms with larger $n$ are suppressed by smaller scales since they have additional factors of $m$, so the largest power of $n$ sets the strong coupling scale.
Raising the strong coupling scale then requires including additional interactions of the form $\partial^k h^3/M_p m^{k-2}$ and  $\partial^k h^4/M_p^2 m^{k-2}$, where the scales suppressing these interactions are fixed by the requirement that they help cancel the existing terms.} If $n\leq 0$ then the amplitudes are perturbatively unitary and can make sense at high energies.

In principle we can solve this problem by writing down all possible cubic and quartic terms in the Lagrangian with arbitrary coefficients, calculating the amplitude, and then choosing the coefficients to minimize $n$.   However, the challenge is that we want to be completely general, allowing for any interactions with an arbitrary (but finite) number of derivatives.  It is therefore advantageous to bypass the Lagrangian and calculate amplitudes directly so that we can avoid redundancies due to field redefinitions and total derivatives.   This allows us to consider general interactions and thus gives a model-independent bound.  

A rough outline of our procedure is the following: first we construct all possible Lorentz-invariant on-shell cubic vertices following the presentation in Ref.~\cite{Costa:2011mg}.  From the point of view of the Lagrangian, these cubic vertices correspond to cubic interactions which cannot be removed by integration by parts or field redefinitions.  For any collection of spins, there are only a finite number of these cubic vertices.
We then use these vertices to construct the exchange amplitudes for four-point tree-level scattering.  Because the cubic vertices are finite in number, there is some maximal growth in energy for the corresponding exchange amplitudes.
Next we construct all possible Lorentz-invariant on-shell quartic vertices that are analytic in momenta.  These correspond to quartic contact terms in the Lagrangian that cannot be removed by integration by parts or field redefinitions.  There are an infinite number of such terms, but they come in a finite number of tensor structures multiplied by polynomials in the Mandelstam invariants.
Adding the exchange and contact amplitudes gives the full tree-level four-particle amplitude consistent with locality.  We then look for the subset of amplitudes that minimize $n$ in Eq.~\eqref{eq:ndef}.  Various technical issues associated with implementing this procedure are discussed in the main text. Some preliminary versions of the results in this paper were presented in a thesis by one of the authors~\cite{Bonifacio:2017iry}.

The rest of the paper is set up in the following way: in Sec.~\ref{sec:construct} we review a procedure for constructing all the cubic and quartic vertices for a given collection of particles. We then work out these vertices in several examples with a single massive particle. In Sec.~\ref{sec:4particle} we define our kinematics and discuss some properties of transversity amplitudes that are needed to simplify our calculation. In Sec.~\ref{sec:method+results} we give a detailed algorithm for finding the most general tree amplitudes with a given high-energy growth. We then present our results for a lower bound on the growth of amplitudes containing a single massive particle. We finish with some discussion in Sec.~\ref{sec:discussion}. The appendices contain some explicit spin-1 tensor structures and amplitudes. 

\textbf{Conventions:} We work in four spacetime dimensions with the mostly plus Lorentzian signature, $\eta_{\mu\nu}={\rm diag}(-1,1,1,1)$. The four-dimensional epsilon symbol is defined with $\varepsilon_{0123}=1$.

\section{Constructing On-Shell Vertices} \label{sec:construct}

The first step in our procedure is to list the possible on-shell cubic and quartic vertices which we will use to construct scattering amplitudes. Constructing such a list amounts to finding all the inequivalent scalar contractions of a given collection of tensors. This problem
also appears in the conformal bootstrap literature, as there is a correspondence between conformal correlators in CFT$_d$ and scattering amplitudes in QFT$_{d+1}$ \cite{Costa:2011mg, Costa:2014rya, Kravchuk:2016qvl}. 
We thus start by reviewing a method for constructing on-shell vertices for a collection of bosonic fields following the approach of Ref.~\cite{Costa:2011mg}. A similar method was used to construct amplitudes for massless particles in Ref.~\cite{Boels:2017gyc}.

We write on-shell vertices in terms of Lorentz-invariant contractions of the polarization vectors and momenta of the external particles. An external particle, labeled by $i$, with spin $l_i$, mass $m_i$, and ingoing momentum $p^i$, is associated with a rank-$l_i$ symmetric, transverse, and traceless polarization tensor $\epsilon^{i}_{\mu_1 \ldots \mu_{l_i}}$,
\begin{align}
\epsilon^{i}_{\mu_1 \ldots \mu_{l_i}}&=\epsilon^{i}_{(\mu_1 \ldots \mu_{l_i})}\,, \\
p_i^{ \mu_1} \epsilon^{i}_{\mu_1 \ldots \mu_{l_i}}&=0, \label{transverse} \\
\eta^{\mu_1 \mu_2}\epsilon^{i}_{\mu_1 \ldots \mu_{l_i}}& =0. \label{traceless} 
\end{align}
To keep track of index contractions, it is convenient to replace each polarization tensor with a product of $l_i$ auxiliary vectors $z^i_{\mu}$ when writing the amplitude,
\be \epsilon^i_{\mu_1 \ldots \mu_{l_i}} \rightarrow z^i_{\mu_1} \ldots z^i_{\mu_{l_i}}.\ee
A vertex coming from a local contact interaction can then be written as a polynomial in the contractions of $z$'s and $p$'s. We denote these contractions by 
\be p_{ij}\equiv p^{i}_{\mu} p^{j, \mu},\ \ z_{ij}\equiv z^{i}_{\mu} z^{j, \mu} ,\ \  zp_{ij}\equiv z^{i}_{\mu} p^{j, \mu}.\ee 
We have the conditions
\bea 
&& z_{ij} =z_{ji} ,\ \  p_{ij} = p_{ji},\label{obvseyms} \\
&& zp_{ii}=0,\ \ \ z_{ii}=0 \label{condeq1d}\, ,\\
&&  p_{ii}=-m_i^2. \label{onshellcondidsx} 
\eea
Here \eqref{obvseyms} comes from the symmetry of the scalar product, \eqref{condeq1d} comes from the transverse, traceless conditions \eqref{transverse} and \eqref{traceless}, and \eqref{onshellcondidsx} comes from the fact that the external momenta are on shell. (Note that repeated instances of the particle labels $i,j,\ldots$ are never summed over.)

Vertices built solely from products of $z_{ij}$, $zp_{ij}$, and $p_{ij}$ will be parity-even. There can also be parity-odd terms that come from contractions involving a single antisymmetric epsilon tensor, $\varepsilon^{\mu_1 \mu_2 \mu_3 \mu_4}$, which we denote by, e.g.
\be \varepsilon (p_1 p_2 z_1 z_2 ) \equiv \varepsilon^{\mu_1 \mu_2 \mu_3 \mu_4} p^1_{\mu_1} p^2_{\mu_2} z^1_{\mu_3} z^2_{\mu_4}.\ee
A theory with a single particle can contain parity-odd cubic terms and still be parity conserving if the particle has odd intrinsic parity and there are no parity-even cubic terms or parity-odd quartic terms. In any case, we do not assume that parity is a symmetry.

If some of the particles interacting at a vertex are identical, then the vertex must be invariant under permutation symmetries of these particles.
There are also dimensionally dependent identities that introduce redundancies between various vertices, which we will need to account for.

\subsection{Cubic vertices}

We start by constructing on-shell cubic vertices. Consider a three-point interaction of particles with spins $(l_1, l_2, l_3)$ and for each $l_i>0$ introduce a null vector $z^i$. Momentum conservation with all momenta incoming gives\footnote{ In general the momenta must be complex for an on-shell cubic amplitude to be nonvanishing. The cubic amplitudes can then be analytically continued to real off-shell momenta to calculate exchange diagrams.}
\be \label{3conservation}
p^{1}+p^{2}+p^{3}=0.
\ee
Eq.~\eqref{3conservation} and $p_{ii}=-m_i^2$ imply that $2 p_{ij} = m_i^2+m_j^2-m_k^2$ for distinct $i, j, k$, so all of the contractions $p_{ij}$ can be written in terms of masses. Contracting \eqref{3conservation} with $z^i$ gives the three relations
\begin{subequations}
\begin{align}
zp_{12}+zp_{13}& =0, \\
zp_{21}+zp_{23}& =0, \\
zp_{31}+zp_{32}& =0 .
\end{align}
\end{subequations}
This means that there are only three independent contractions $zp_{ij}$, rather than six. From the point of view of the Lagrangian, this reduction comes from the freedom to integrate by parts.

\subsubsection{Parity even}
In total, there are therefore six independent Lorentz scalars that can be used to construct the on-shell parity-even cubic vertices, which can be taken to be 
\be z_{12}, \ z_{13},\ z_{23},\ zp_{12},\ zp_{23}, \ zp_{31}.
\ee
Each $z_i$ must appear $l_i$ times in the cubic vertex, since each polarization vector appears once. The amplitude is thus a linear combination of terms of the form
\be \label{eq:cubicamp}
z_{12}^{n_{12}} z_{13}^{n_{13}} z_{23}^{n_{23}} zp_{12}^{m_{12}} zp_{23}^{m_{23}} zp_{31}^{m_{31}} ,
\ee
where the exponents $n_{ij}$ and $m_{ij}$ are non-negative integers satisfying
\begin{subequations}
\label{eq:cubiccond}
\begin{align}
n_{12}+n_{13}+m_{12}& =l_1, \\
n_{12}+n_{23}+m_{23}& =l_2, \\
n_{13}+n_{23}+m_{31}& = l_3.
\end{align}
\end{subequations}
There are a finite number of solutions to Eqs.~\eqref{eq:cubiccond} for a given triplet of spins. 
For identical particles we must additionally look for combinations of the vertices that are invariant under the action of the symmetric group $S_3$ of particle permutations.

\subsubsection{Parity odd}

There can also be parity-odd cubic vertices containing one of the terms $\varepsilon (p_1 p_2 z_1 z_2 )$, $\varepsilon (p_1 p_2 z_1 z_3 )$, $\varepsilon (p_1 p_2 z_2 z_3 )$, $\varepsilon (p_1 z_1 z_2 z_3 )$, or $\varepsilon (p_2 z_1 z_2 z_3 )$. These may have to be multiplied by parity-even structures so that the result contains enough $z$'s. The general parity-odd cubic vertex is a linear combination of terms of the form
\be
\varepsilon (z^{\eta_1}_1 z^{\eta_2}_2 z^{\eta_3}_3 p^{\eta_4}_1 p^{\eta_5}_2) z_{12}^{\tilde{n}_{12}} z_{13}^{\tilde{n}_{13}} z_{23}^{\tilde{n}_{23}} zp_{12}^{\tilde{m}_{12}} zp_{23}^{\tilde{m}_{23}} zp_{31}^{\tilde{m}_{31}},
\ee
where $\eta_i \in \{0, 1\}$ and $\tilde{n}_{ij}, \tilde{m}_{ij}$ are non-negative integers satisfying
\begin{subequations}
\label{eq:cubiccondodd}
\begin{align}
\tilde{n}_{12}+\tilde{n}_{13}+\tilde{m}_{12}+\eta_1& =l_1, \\
\tilde{n}_{12}+\tilde{n}_{23}+\tilde{m}_{23}+\eta_2& =l_2, \\
\tilde{n}_{13}+\tilde{n}_{23}+\tilde{m}_{31}+\eta_3& = l_3, \\
\eta_1+\eta_2+\eta_3+\eta_4+\eta_5&=4.
\end{align}
\end{subequations}
The number of solutions to Eqs. \eqref{eq:cubiccondodd} is also finite, 
and for identical particles we must again look for $S_3$-invariant combinations of these structures.

\subsubsection{Dimensionally dependent identities}
In four dimensions there can be redundancies amongst the cubic vertices due to dimensionally dependent (Gram) identities. These can be written as contractions of the five-dimensional epsilon tensor. In the parity-even case we have the four-dimensional identity
\be \label{eq:cubicevenDDI}
\varepsilon(p_1 p_2 z_1 z_2 z_3) \varepsilon(p_1 p_2 z_1 z_2 z_3)=0,
\ee
where the epsilons taken together should be understood here in terms of the generalized Kronecker delta tensor $\delta^{[\mu_1}_{\nu_1} \ldots \delta^{\mu_5]}_{\nu_5}$.
This identity eliminates one combination of the cubic vertices for any collection of spin-2 particles \cite{Bonifacio:2017nnt}.

In the parity-odd case we can obtain four-dimensional identities by contracting a single five-dimensional epsilon tensor with linearly dependent five-vectors
\be \label{eq:cubicoddDDI}
\varepsilon \left(P_1 P_2 Z_1 Z_2 Z_3 \right) = 0,
\ee
where
\be
Z^{A}_{1} = \left( \begin{matrix} A_1 \\ z_{1}^{\mu} \end{matrix} \right), \, Z^{A}_{2} = \left( \begin{matrix} A_2 \\ z_{2}^{\mu} \end{matrix} \right), \, P^{A}_{1} = \left( \begin{matrix} A_3 \\ p_{1}^{\mu} \end{matrix} \right), \, P^{A}_{2} = \left( \begin{matrix} A_4 \\ p_{2}^{\mu} \end{matrix} \right), \, Z^{A}_{3} = \left( \begin{matrix} \sum_{i=1}^4 \alpha_i A_i  \\ z_{3}^{\mu} \end{matrix} \right).
\ee
Here $A_1$, $A_2$, $A_3$, $A_4$ are any scalars and  $\alpha_i$ are defined by $z_3 = \alpha_1 z_1 + \alpha_2 z_2 + \alpha_3 p_1 + \alpha_4 p_2$ \cite{Costa:2011mg}. For example, choosing $\{A_1, A_2, A_3, A_4 \} = \{ z_{1j}, z_{2j}, zp_{j1}, zp_{j2} \}$ for $j=1,2,3$ and $\{A_1, A_2, A_3, A_4 \} = \{ zp_{1j}, zp_{2j}, p_{j1}, p_{j2} \}$ for $j=1,2$ gives the five four-dimensional identities
\begin{subequations} \label{eq:parityoddid}
\begin{align}
& z_{13} \varepsilon(p_1p_2 z_1 z_2)-z_{12} \varepsilon(p_1p_2 z_1 z_3)-zp_{12} \varepsilon(p_1 z_1 z_2 z_3)=0, \\
& z_{12} \varepsilon(p_1p_2 z_2 z_3)+z_{23} \varepsilon(p_1p_2 z_1 z_2)-zp_{23} \varepsilon(p_2 z_1 z_2 z_3)=0, \\
& z_{13} \varepsilon(p_1p_2 z_2 z_3)-z_{23} \varepsilon(p_1p_2 z_1 z_3)+zp_{31}\left(  \varepsilon(p_1 z_1 z_2 z_3)+\varepsilon(p_2 z_1 z_2 z_3)\right)=0, \\
& zp_{23} \varepsilon(p_1p_2 z_1z_3)+ zp_{31} \varepsilon(p_1p_2 z_1z_2)+p_{11} \varepsilon(p_2 z_1 z_2z_3)-p_{12} \varepsilon(p_1 z_1 z_2z_3) =0, \label{eq:oddidentity4}\\
& zp_{12} \varepsilon(p_1p_2 z_2 z_3)- zp_{31} \varepsilon(p_1p_2 z_1z_2)+p_{12} \varepsilon(p_2 z_1 z_2z_3)-p_{22} \varepsilon(p_1 z_1 z_2z_3) =0.\label{eq:oddidentity5}
\end{align}
\end{subequations}

To obtain a structure with enough $z$'s from the identities \eqref{eq:cubicoddDDI} or \eqref{eq:cubicevenDDI} we may have to multiple by additional $z_{ij}$ or $zp_{ij}$ terms. There are more ways to do this in the parity-odd case since the identities contain fewer $z$'s to begin with. Together with the freedom to choose the scalars $A_i$, this results in many more parity-odd identities than parity-even ones. This will be reflected below in the larger redundancy of parity-odd cubic and quartic vertices as compared to parity-even vertices.  Accounting for these dimensionally dependent identities, the independent tensor structures for different collections of particles in various spacetime dimensions have been nicely enumerated using the representation theory of stabilizer groups by Kravchuk and Simmons-Duffin \cite{Kravchuk:2016qvl}.\footnote{Another approach to counting independent operators is through Hilbert series~\cite{Henning:2017fpj}.} For the case of identical particles in four dimensions with integer spin $l$, their results for the number of parity-even and parity-odd cubic structures can be written as 
\be \label{eq:numcubiceven}
N_\text{cubic}(l) =\floor*{\frac{l}{2}}^2 +\frac{\left(1+(-1)^l\right)\left(l+1\right)}{2}
\ee
and
\be \label{eq:numcubicodd}
\tilde{N}_\text{cubic}(l) =\floor*{\frac{l}{2}}\left( \floor*{\frac{l}{2} }+1\right),
\ee
where $\floor*{\cdot}$ is the floor function.

\subsection{Quartic vertices}

Now consider a four-point interaction of particles with spins $(l_1, l_2, l_3, l_4)$. Momentum conservation with all momenta incoming gives
\be \label{4conservation}
p^{1}+p^{2}+p^{3}+p^4=0.
\ee
Using the conditions \eqref{4conservation} and $p_{ii}=-m_i^2$, it can be shown that there are only two independent contractions $p_{ij}$ for $i \ne j$, which we take to be $p_{12}$ and $p_{13}$. These correspond to the two independent Mandelstam variables $s$ and $t$ for $2 \rightarrow 2$ scattering, which for all momenta incoming are defined by
\be s=-2p_{12}+m_1^2+m_2^2,\ \ \ t=-2p_{13}+m_1^2+m_3^2. \label{stmandelse}\ee
 Contracting Eq.~\eqref{4conservation} with $z^i$ gives the four relations
\begin{subequations}
\begin{align}
zp_{12}+zp_{13}+zp_{14} & =0, \\
zp_{21}+zp_{23}+zp_{24}& =0, \\
zp_{31}+zp_{32}+zp_{34}& =0, \\
zp_{41}+zp_{42}+zp_{43}& =0. 
\end{align}
\end{subequations} 
This means that there are only eight independent contractions of the form $zp_{ij}$, rather than 12. There are also six contractions of the form $z_{ij}$. 

\subsubsection{Parity even}
In total, there are 16 independent parity-even Lorentz scalars. The parity-even quartic vertices are thus built from tensor structures of the form 
\be \label{eq:quarticamp}
 z_{12}^{n_{12}} z_{13}^{n_{13}} z_{14}^{n_{14}} z_{23}^{n_{23}} z_{24}^{n_{24}} z_{34}^{n_{34}} zp_{13}^{m_{13}} zp_{14}^{m_{14}} zp_{21}^{m_{21}} zp_{24}^{m_{24}} zp_{31}^{m_{31}} zp_{32}^{m_{32}} zp_{42}^{m_{42}} zp_{43}^{m_{43}},
\ee
where the exponents $n_{ij}$ and $m_{ij}$ are non-negative integers satisfying 
\begin{subequations}
\label{eq:quarticcond}
\begin{align}
n_{12}+n_{13}+n_{14}+m_{13}+m_{14} & =l_1, \\
n_{12}+n_{23}+n_{24}+m_{21}+m_{24} & =l_2, \\
n_{13}+n_{23}+n_{34}+m_{31}+m_{32} & =l_3, \\
n_{14}+n_{24}+n_{34}+m_{42}+m_{43} & =l_4.
\end{align}
\end{subequations}
There are a finite number of solutions to Eqs.~\eqref{eq:quarticcond}.
The corresponding tensor structures encode the different ways of contracting the polarization tensors. Each tensor structure can also be multiplied by a function $f(s,t)$ of the Mandelstam variables, reflecting the fact that there are on-shell quartic interactions with arbitrarily many derivatives. (In a CFT this corresponds to the freedom to multiply a four-point conformal correlator by a function of the conformal cross ratios.) When a general four-point amplitude is expanded in terms of these quartic tensor structures, the coefficient $f(s,t)$ can have nonanalytic pieces due to contributions from, e.g. loops and exchange diagrams. However, at tree level the contributions to $f(s,t)$ from local contact interactions are polynomial. 

As in the cubic case, permutation symmetries impose extra constraints on the quartic vertices. When the four external particles are identical we should look for combinations of the vertices that are invariant under the symmetric group $S_4$ of particle permutations. However, some of these permutations will interchange the Mandelstam variables, which means that the functions appearing in front of the different tensor structures satisfy additional crossing relations. 
Instead of imposing $S_4$ invariance and solving the crossing relations in terms of the functions $f_{\mathcal{I}}(s,t)$, our approach is to deal with tensor structures that are invariant under the subgroup of permutations that do not affect the Mandelstam variables, which are called kinematic permutations~\cite{Kravchuk:2016qvl}. The remaining permutation symmetries then correspond to crossing symmetries of the scattering amplitudes, which we later impose as an additional constraint. 

For the case of four identical particles that we consider,  the group of kinematic permutations has four elements and is isomorphic to $\mathbb{Z}_2^2$~\cite{Kravchuk:2016qvl}. In cycle notation it is given by 
\be \label{eq:kinperms}
\Pi^{\rm kin} = \{ {\cal I}, (12)(34), (13)(24), (14)(23) \}  \cong \mathbb{Z}_2^2, 
\ee
where ${\cal I}$ is the identity element.  We denote the parity-even $\mathbb{Z}_2^2$-invariant tensor structures by $\mathbb{T}_I$. The most general parity-even four-point amplitude for identical particles coming from tree-level contact diagrams is then given by
\be \label{eq:genquartic}
\mathcal{A}_{\rm contact} = \sum_{I} f_I(s,t) \mathbb{T}_I,
\ee
where $f_I(s,t)$ are arbitrary polynomials (up to crossing constraints) and the sum runs over all of the $\mathbb{Z}_2^2$-invariant tensor structures.\footnote{In \eqref{eq:genquartic} we have assumed that the $\mathbb{Z}_2^2$-invariant amplitudes can be written in terms of $\mathbb{Z}_2^2$-invariant tensor structures. We could also add combinations of unsymmetrized tensor structures that vanish due to dimensionally dependent identities, but these would be projected away when we substitute four-dimensional kinematics in Sec.~\ref{ssec:detailsofmethod}, so we can ignore them without loss of generality.}

\subsubsection{Parity odd}

There are also parity-odd quartic structures. These contain one of the 35 independent contractions of the form $\varepsilon (p_1 p_2 p_3 z_i )$, $\varepsilon (p_i p_j z_k z_l )$, $\varepsilon (p_i z_j z_k z_l )$, or $\varepsilon (z_1 z_2 z_3 z_4 )$. The general parity-odd quartic structure is made from terms of the form
\be \label{eq:quarticampodd}
\varepsilon (z^{\eta_1}_1 z^{\eta_2}_2 z^{\eta_3}_3 z^{\eta_4}_4 p^{\eta_5}_1 p^{\eta_6}_2 p^{\eta_7}_3) z_{12}^{\tilde{n}_{12}} z_{13}^{\tilde{n}_{13}} z_{14}^{\tilde{n}_{14}} z_{23}^{\tilde{n}_{23}} z_{24}^{\tilde{n}_{24}} z_{34}^{\tilde{n}_{34}} zp_{13}^{\tilde{m}_{13}} zp_{14}^{\tilde{m}_{14}} zp_{21}^{\tilde{m}_{21}} zp_{24}^{\tilde{m}_{24}} zp_{31}^{\tilde{m}_{31}} zp_{32}^{\tilde{m}_{32}} zp_{42}^{\tilde{m}_{42}} zp_{43}^{\tilde{m}_{43}},
\ee
where $\eta_i \in \{0, 1\}$ and $\tilde{n}_{ij}, \tilde{m}_{ij}$ are non-negative integers satisfying 
\begin{subequations}
\label{eq:quarticcondodd}
\begin{align}
\tilde{n}_{12}+\tilde{n}_{13}+\tilde{n}_{14}+\tilde{m}_{13}+\tilde{m}_{14} +\eta_1& =l_1, \\
\tilde{n}_{12}+\tilde{n}_{23}+\tilde{n}_{24}+\tilde{m}_{21}+\tilde{m}_{24}+\eta_2 & =l_2, \\
\tilde{n}_{13}+\tilde{n}_{23}+\tilde{n}_{34}+\tilde{m}_{31}+\tilde{m}_{32} +\eta_3& =l_3, \\
\tilde{n}_{14}+\tilde{n}_{24}+\tilde{n}_{34}+\tilde{m}_{42}+\tilde{m}_{43} +\eta_4& =l_4, \\
\eta_1+\eta_2+\eta_3+\eta_4+\eta_5+\eta_6 + \eta_7&= 4.
\end{align}
\end{subequations}
We denote the parity-odd $\mathbb{Z}_2^2$-invariant tensor structures by  $\tilde{\mathbb{T}}_{\tilde{I}}$. The most general parity-odd four-point amplitude for identical particles coming from tree-level contact diagrams is thus given by
\be \label{eq:genquarticodd}
\tilde{\mathcal{A}}_{\rm contact} = \sum_{\tilde{I}} \tilde{f}_{\tilde{I}}(s,t) \tilde{\mathbb{T}}_{\tilde{I}},
\ee
where $\tilde{f}_{\tilde{I}}(s,t)$ are arbitrary polynomials (up to crossing constraints) and the sum runs over all of the $\mathbb{Z}_2^2$-invariant tensor structures.

\subsubsection{Dimensionally dependent identities}

As for cubic vertices, there are redundant combinations of quartic structures in four dimensions due to dimensionally dependent identities. These identities can be constructed from contractions of five-dimensional epsilon tensors, as in Eqs.~\eqref{eq:cubicevenDDI} and \eqref{eq:cubicoddDDI}.  There are now seven vectors available for contraction, $\{z_1, z_2, z_3, z_4, p_1, p_2, p_3\}$, and additional $z$'s are required to form a quartic structure, so there are more identities than in the cubic case.  

Due to these identities, the sets of tensor structures obtained by solving Eqs.~\eqref{eq:quarticcond} and \eqref{eq:quarticcondodd} are generally not independent.  They satisfy relations of the form
\begin{align} \label{eq:identities}
\sum_{I} c^{(n)}_I(s,t) \mathbb{T}_{I}=0, \quad  \sum_{\tilde{I}} \tilde{c}^{(\tilde{n})}_{\tilde{I}}(s,t) \tilde{\mathbb{T}}_{\tilde{I}}=0,
\end{align}
where $c^{(n)}_I(s,t)$ and $\tilde{c}^{(\tilde{n})}_{\tilde{I}}(s,t)$ are polynomials in $s$ and $t$.\footnote{Mathematically, before imposing permutation symmetries the quartic contact amplitudes are elements of a module over the ring of polynomials in $s$ and $t$ with real coefficients. This module is generated by the set of tensor structures and the $\mathbb{Z}_2^2$-invariant amplitudes form a submodule. Unlike vector spaces, modules do not always have a basis due to the possibility of nontrivial relations amongst the generators as in \eqref{eq:identities}, which are called syzygies. Modules that have a basis are called free modules.} 
Ideally we would use these relations to eliminate redundant tensor structures, e.g. by writing
\be
\mathbb{T}_{I_0} = -\frac{1}{c^{(n_0)}_{I_0}(s,t) } \sum_{I \neq I_0} c^{(n_0)}_I(s,t) \mathbb{T}_{I}
\ee
in the expansion \eqref{eq:genquartic} and then absorbing this by redefining the coefficients,
\be
f_{I}(s,t) \rightarrow f_{I}(s,t) +\frac{f_{I_0}(s,t) c^{(n_0)}_{I}(s,t)}{c^{(n_0)}_{I_0}(s,t)},
\ee
for $I \neq I_0$. However,
if $c^{(n_0)}_{I_0}(s,t)$ is not a constant then this would result in the new coefficients becoming rational functions if they started out as polynomials. This creates a tension between writing contact amplitudes in terms of independent tensor structures and manifesting locality through polynomial coefficients. When $c^{(n_0)}_{I_0}(s,t)$ is a nonzero constant, then it is possible to eliminate the structure $\mathbb{T}_{I_0}$ and still have polynomial coefficients, although achieving this may require redefining some structures. If it is possible to eliminate all syzygies in this way then the resulting module is free. In practice it can be difficult to find a basis, so it is sometimes easier to use tensor structures that are not independent. 

The numbers of independent parity-even and parity-odd $\mathbb{Z}_2^2$-invariant quartic structures for bosonic spin-$l$ particles are given by
\begin{align}
N_\text{quartic}(l) &= 1+2 l(1+l)(2+l+l^2), \label{eq:Neven} \\
\tilde{N}_\text{quartic}(l) &= l(1+l)(1+2l+2l^2) \label{eq:Nodd},
\end{align}
respectively~\cite{Kravchuk:2016qvl}. So a basis of these tensor structures, if it exists, has $N_\text{quartic}(l)$ or $\tilde{N}_\text{quartic}(l)$ elements.

One way to find syzygies is to search systematically for combinations of tensor structures amongst a general superposition of the dimensionally dependent identities. Alternatively, we can evaluate the tensor structures using explicit four-dimensional kinematics and search for combinations of the form \eqref{eq:identities} that vanish when written in terms of $s$ and $t$. A method to directly construct a basis is the conformal/scattering frame approach of \cite{Kravchuk:2016qvl}, as employed for the 3$d$ stress-tensor bootstrap in \cite{Dymarsky:2017yzx}. Further, a good check for whether a set of $n$-point tensor structures is independent is to calculate the `$P$-matrix'~\cite{Boels:2017gyc}, defined by
\be \label{eq:Pmatrix}
P_{I J} = \sum_{\tau_1, \ldots, \tau_n} T^{* \tau_1\ldots \tau_n}_I T^{\tau_1\ldots \tau_n}_J,
\ee
where the sum runs over a basis of polarizations for each external particle. This can be written in terms of Mandelstam variables using the on-shell completeness relation
\be
\sum_{\tau_i} \epsilon^{(\tau_i)}_{\mu_1 \ldots \mu_{l_i}}\epsilon^{* (\tau_i)}_{\nu_1 \ldots \nu_{l_i}} = \Pi_{\mu_1 \ldots \mu_{l_i}, \nu_1 \ldots \nu_{l_i}},
\ee
where $\epsilon^{(\tau_i)}_{\mu_1 \ldots \mu_{l_i}}$ is a basis of polarization tensors for particle $i$ and $\Pi_{\mu_1 \ldots \mu_{l_i}, \nu_1 \ldots \nu_{l_i}}$ is the spin-$l_i$ projector (see Sec.~\ref{sec:4particle}). If the determinant of $P_{I J}$ is nonzero then the tensor structures are independent.

\subsection{Examples} \label{sec:amplitudes} 

We now apply this formalism to explicitly construct all the cubic and quartic vertices for identical self-interacting massive fields with integer spin $\leq 2$ in four dimensions. We also discuss the zero-derivative quartic vertices for higher-spin particles.

\subsubsection{Spin 0}

\subsubsection*{Cubic vertices}
For scalar fields the only solution to the cubic equations \eqref{eq:cubiccond} is $n_{ij}=m_{ij}=0$, which gives a constant, $\lambda$. 
For a single self-interacting scalar, this corresponds to the interaction
$\lambda \phi^3/6$.  Any other cubic interactions in the Lagrangian can be written in terms of this one and higher-order interactions after integrating by parts and redefining fields.  There are no parity-odd cubic vertices.

\subsubsection*{Quartic vertices}
At the quartic level there is only one tensor structure for a scalar, the constant vertex $\mathbb{T}_1 = 1$, so the quartic contact amplitude is given by
\be \label{eq:scalarquartic}
\mathcal{A}_{\rm contact} = f_1(s,t),
\ee
where the function $f_1(s,t)$ is an arbitrary polynomial (up to crossing constraints). There are no parity-odd quartic structures.

\subsubsection{Spin 1}

\subsubsection*{Cubic vertices}
The cubic vertices for spin 1 are found by solving Eq.~\eqref{eq:cubiccond} with $l_i =1$. There are four solutions, giving the structures
\be
z_{12} zp_{31}, \; z_{13} zp_{23}, \; z_{23} zp_{12}, \; zp_{12} zp_{23} zp_{31}.
\ee
No combination of these is invariant under all permutations of the particles, so there are no on-shell parity-even cubic amplitudes for a single vector.\footnote{This implies that there can be no cubic photon interactions.  There are cubic amplitudes for multiple vectors, e.g. the Yang-Mills cubic vertex 
\be
f^{abc} \left(z_{12}^{ab} zp_{31}^c+z_{13}^{ac}zp_{23}^b + z_{23}^{bc} zp_{12}^a \right),
\ee 
where $f^{abc}$ is an antisymmetric structure constant.} There are also no parity-odd cubic terms for a single vector; the five parity-odd structures obtained by solving \eqref{eq:cubiccondodd} are
\be
\varepsilon (p_1 p_2 z_1 z_2 ) zp_{31}, \; \varepsilon (p_1 p_2 z_1 z_3 ) zp_{23}, \; \varepsilon (p_1 p_2 z_2 z_3) zp_{12}, \; \varepsilon (p_1 z_1 z_2 z_3 ), \; \varepsilon (p_2 z_1 z_2 z_3),
\ee
but no combination of these is invariant under the symmetric group $S_3$.

\subsubsection*{Quartic vertices}
To find the parity-even quartic vertices we must solve Eq.~\eqref{eq:quarticcond} with $l_i=1$. Imposing the $\mathbb{Z}_2^2$ permutation symmetry leaves 19 structures $\mathbb{T}_I$, although by Eq.~\eqref{eq:Neven} only 17 of these are independent. We can eliminate two of the structures using the following dimensionally dependent identities:
\begin{align}
 \varepsilon \left(p_1 p_2 p_3 z_1 z_2 \right) \varepsilon \left(p_1 p_2 p_3 z_3 z_4 \right) =0,\\
 \varepsilon \left(p_1 p_2 p_3 z_1 z_3 \right) \varepsilon \left(p_1 p_2 p_3 z_2 z_4 \right) =0.
\end{align} 
These are also the only independent identities that have the form of a parity-even spin-1 quartic vertex, which explains the initial amount of redundancy. We list explicitly our basis of 17 parity-even structures in Appendix \ref{app:structures}.

We can similarly find the parity-odd quartic structures by solving Eq.~\eqref{eq:quarticampodd} with $l_i=1$. After imposing the $\mathbb{Z}_2^2$ symmetry we find 48 structures, but by \eqref{eq:Nodd} only ten of these are independent. Despite this large redundancy, we can find a basis of structures while preserving manifest locality (keeping the coefficients $\tilde{f}_{\tilde{I}}(s,t)$ as polynomials) by using dimensionally dependent identities. Our basis of ten parity-odd structures is given explicitly in Appendix \ref{app:structures}. We have checked that the tensor structures in these parity-even and parity-odd bases are independent by evaluating the determinant of the matrix $P_{I J}$ defined in Eq.~\eqref{eq:Pmatrix}.

To summarize, the general spin-1 tree-level quartic contact amplitude can be written in the form
\be \label{eq:spin1quartic}
\mathcal{A}_{\rm contact} +\tilde{\mathcal{A}}_{\rm contact}  = \sum_{I=1}^{17} f_I(s,t) \mathbb{T}_I+\sum_{\tilde{I}=1}^{10} \tilde{f}_{\tilde{I}}(s,t) \tilde{\mathbb{T}}_{\tilde{I}},
\ee
where $\mathbb{T}_I, \tilde{\mathbb{T}}_{\tilde{I}}$ are the tensor structures listed explicitly in Appendix \ref{app:structures} and $f_I(s,t), \tilde{f}_{\tilde{I}}(s,t)$ are polynomials in the Mandelstam variables.

\subsubsection{Spin 2}

\subsubsection*{Cubic vertices}
The parity-even cubic vertices for a spin-2 field are found by solving Eq.~\eqref{eq:cubiccond} with $l_i =2$. There are 11 solutions in total. 
Five combinations of the corresponding tensor structures are invariant under permuting the particles, corresponding to the cubic vertices for identical particles. These cubic vertices are given by
\begin{subequations}\label{eq:Astructures}
\begin{align}
\mathcal{A}_1 = & z_{12} z_{13} z_{23}, \label{eq:spin2cubic1} \\
\mathcal{A}_2 = & z_{23}^2 zp_{12}^2+z_{13}^2 zp_{23}^2+z_{12}^2 zp_{31}^2, \label{eq:spin2cubicGR}\\
\mathcal{A}_3 = & z_{13} z_{23}zp_{12}zp_{23}+z_{12}z_{23}zp_{12}zp_{31}+z_{12}z_{13}zp_{23} zp_{31}, \\
\mathcal{A}_4 = &zp_{12} zp_{23} zp_{31} \left( z_{12}zp_{31}+z_{23} zp_{12} +z_{13} zp_{23} \right),\\
\mathcal{A}_5 = & zp_{12}^2 zp_{23}^2 zp_{31}^2.
\end{align}
\end{subequations}
The dimensionally dependent identity \eqref{eq:cubicevenDDI} gives the relation
\be
4 \mathcal{A}_4 - 2 m^2 \left( \mathcal{A}_2+ \mathcal{A}_3 \right) +3m^4 \mathcal{A}_1=0,
\ee
so we can eliminate the structure $\mathcal{A}_4$ in four dimensions by writing it in terms of the others.  This leaves four independent structures, in agreement with \eqref{eq:numcubiceven}. 
The most general parity-even cubic vertex is given by
\be
g_1 \mathcal{A}_1+g_2 \mathcal{A}_2+g_3 \mathcal{A}_3+g_5 \mathcal{A}_5,
\ee
where $g_n$ are cubic coupling constants. In dRGT massive gravity the two-derivative cubic couplings satisfy $g_3/g_2=2$, which is a signature of the Einstein-Hilbert kinetic term, whereas in the pseudolinear theory they satisfy $g_3/g_2=1$. Both of these theories also have $g_5=0$ and $g_1$ as a free parameter. See~\cite{Hinterbichler:2017qyt} for a detailed comparison of the structures \eqref{eq:Astructures} to those generated by familiar Lagrangian interactions.  

Now consider parity-odd cubic vertices. Solving Eq.~\eqref{eq:cubiccondodd} and symmetrizing gives the following five structures:
\begin{subequations}
\begin{align}
\mathcal{B}_1 =& z_{13} z_{23} \varepsilon (p_{1} p_2 z_{1} z_2 )-z_{12} z_{23} \varepsilon (p_{1} p_2 z_{1} z_3)+z_{12} z_{13} \varepsilon (p_{1} p_2 z_{2} z_3),  \\
\mathcal{B}_2 =&  z_{12} zp_{31}\varepsilon (p_{2}z_{1} z_2 z_3 )-z_{13} zp_{23}\varepsilon (p_{2}z_{1} z_2 z_3 )+\left( z_{12}zp_{31}-z_{23}zp_{12}\right)\varepsilon (p_{1}z_{1} z_2 z_3 ),\\
\mathcal{B}_3 =& z_{12} zp_{31}^2 \varepsilon (p_{1} p_2 z_{1} z_2 )+ z_{23} zp_{12}^2 \varepsilon (p_{1} p_2 z_{2} z_3 )- z_{13} zp_{23}^2 \varepsilon (p_{1} p_2 z_{1} z_3), \\
\mathcal{B}_4 =&zp_{31}\left(z_{23} zp_{12}+z_{13}zp_{23} \right) \varepsilon (p_{1} p_2 z_{1} z_2 ) +zp_{12}\left(z_{13} zp_{23}+z_{12}zp_{31} \right) \varepsilon (p_{1} p_2 z_{2} z_3 ) \nonumber \\&-zp_{23}\left(z_{23} zp_{12}+z_{12}zp_{31} \right) \varepsilon (p_{1} p_2 z_{1} z_3 )  ,\\
\mathcal{B}_5 =&zp_{12} zp_{23} zp_{31} \left( zp_{31} \varepsilon ( p_1 p_2 z_1 z_2 ) - zp_{23} \varepsilon ( p_1 p_2 z_1 z_3)+ zp_{12} \varepsilon ( p_1 p_2 z_2 z_3)\right) \label{eq:oddRiemann}.
\end{align}
\end{subequations}
Using the four-dimensional identities \eqref{eq:parityoddid} we get the constraints
\begin{subequations}
\begin{align}
\mathcal{B}_4 & =0, \\
2\mathcal{B}_1+\mathcal{B}_2 &=0,\\
3m^2 \mathcal{B}_1+2\mathcal{B}_3 &=0.
\end{align}
\end{subequations}
This means that only two of the structures are independent, in agreement with \eqref{eq:numcubicodd}. We take $\mathcal{B}_1$ and $\mathcal{B}_5$ as a basis and write the most general parity-odd cubic vertex as
\be
\tilde{g}_1 \mathcal{B}_1+\tilde{g}_5 \mathcal{B}_5,
\ee
where $\tilde{g}_{\tilde{n}}$ are cubic coupling constants.

The vertex $\mathcal{B}_5$ is generated by the cubic part of the parity-odd Riemann-cubed interaction 
\be
\sqrt{-g}\left(\widetilde{R}_{\mu \nu}{}^{ \alpha \beta} R_{\alpha \beta}{}^{\lambda \rho} R_{\lambda \rho}{}^{\mu \nu}\right),
\ee
where $\widetilde{R}_{\mu \nu}{}^{ \alpha \beta} \equiv \varepsilon_{\mu \nu \lambda \rho}{R}^{\lambda \rho \alpha \beta}$.
If the external particles are massless, then $\mathcal{B}_5$ is gauge invariant and using the identities \eqref{eq:oddidentity4} and \eqref{eq:oddidentity5} it can be written more simply as
\be
3 zp_{12} zp_{23} zp_{31}^2 \varepsilon ( p_1 p_2 z_1 z_2 ), 
\ee
which agrees with the form of this structure presented in \cite{Camanho:2014apa}. 
The vertex $\mathcal{B}_1 $ is generated by the Lagrangian interaction
\be \label{eq:parityoddgraviton}
\varepsilon^{\mu \nu \lambda \rho} \partial_{\mu} h_{\nu \alpha} \partial_{\lambda} h_{\rho \beta} h^{\alpha \beta}.
\ee
We do not know of any theory that utilizes this vertex. It has two derivatives, is not gauge invariant, and exists only in four dimensions, so it is a good candidate interaction for a four-dimensional parity-odd theory of a massive spin-2 particle.\footnote{There are no parity-odd cubic vertices for identical spin-2 particles in five dimensions since there is no $S_3$-invariant combination of structures containing $\varepsilon(p_1 p_2 z_1 z_2 z_3)$. In three dimensions there is a well-known parity-odd theory of massive gravity, namely topological massive gravity \cite{Deser:1981wh}. A parity-odd interaction for massive gravity in the vielbein formulation was discussed in~\cite{Goon:2014paa}.}

\subsubsection*{Quartic vertices}
To find the parity-even quartic vertices we must solve Eq.~\eqref{eq:quarticcond} with $l_i=2$.  After imposing the kinematic permutations there are 201 parity-even structures, but only 97 of these are independent by Eq.~\eqref{eq:Neven}. 
We did not find a basis of these structures, so we work with structures that are not independent. 

We can similarly find all the quartic parity-odd structures by solving \eqref{eq:quarticcondodd} with $l_i=2$. After imposing the kinematic permutations there are 1266 parity-odd structures, but only 78 of these are independent by Eq.~\eqref{eq:Nodd}. This huge redundancy comes from the large number of dimensionally dependent identities involving parity-odd structures. A simple way to reduce the number of structures is to find relations of the form \eqref{eq:identities} where \textit{all} of the coefficients $\tilde{c}^{(\tilde{n})}_{\tilde{I}}(s,t)$ are constants. There are 804 relations of this form, so this reduces the number of structures to 462. This is enough of a reduction to make the calculation of the general amplitude feasible, so we do not attempt to find a smaller set. 

Overall, the general tree-level quartic contact amplitude for a self-interacting spin-2 particle is given by
\be
\mathcal{A}_{\rm contact} +\tilde{\mathcal{A}}_{\rm contact} = \sum_{I=1}^{201} f_I(s,t) \mathbb{T}_I+\sum_{\tilde{I}=1}^{462} \tilde{f}_{\tilde{I}}(s,t) \tilde{\mathbb{T}}_{\tilde{I}},
\ee
where $f_I(s,t)$ and $\tilde{f}_{\tilde{I}}(s,t)$ are polynomials. The spin-2 tensor structures are too numerous for us to list them explicitly, but they can be generated using the \texttt{Mathematica} notebook attached as an ancillary file.

\subsubsection{Spin $l$}
We could repeat the previous construction for higher spins, but actually calculating the general amplitude becomes difficult as the number of tensor structures increases. The general case for spins greater than 2 is beyond the scope of this paper, but we instead consider the simpler problem of finding all zero-derivative quartic contact structures.  It also seems plausible that these suffice to determine a lower bound on the growth of the four-point amplitude. 

When constructing a local Lagrangian for massive higher-spin particles, it is necessary to introduce auxiliary fields to help enforce the required on-shell constraints \cite{Singh:1974qz}. These fields vanish on shell, so an advantage of working directly with on-shell amplitudes is that we avoid having to introduce auxiliary fields.  

\subsubsection*{Zero-derivative quartic vertices}
To find the zero-derivative quartic vertices for a single massive spin-$l$ particle we must solve
\be \label{eq:zerodeq}
n_{12}+n_{13}+n_{14}=l, \quad n_{23}=n_{14}, \quad  n_{24} = n_{13}, \quad n_{34} =n_{12}.
\ee
There are $(l+1)(l+2)/2$ solutions to these equations, so this is the number of structures before symmetrizing. 

We next impose a permutation symmetry on these structures. In this case we can impose the full $S_4$ permutation symmetry, rather than just the $\mathbb{Z}_2^2$ subgroup. This is because we ignore derivative interactions, so the coefficients $f_I(s,t)$ are just constants. It is therefore straightforward to find $S_4$-invariant combinations of vertices, which means that the resulting amplitudes will automatically be crossing symmetric. 

Using Burnside's lemma we get that the number of $S_4$-invariant structures is
 \be \label{eq:zerodnumber}
\frac{6+3l+l^2+6 \floor{\frac{2+l}{2}}+4\floor{\frac{1}{2} \cos \left(\frac{2 \pi l}{3}\right)} }{12} = \left[ \frac{(l+3)^2}{12}\right],
\ee
where $[ \cdot ]$ is the nearest integer function. It is straightforward to construct these vertices for any given $l$ and they are independent in four dimensions. 

We can also write down the corresponding on-shell Lagrangian interactions: schematically they are of the form
\be
\phi_{\mu_1\ldots \mu_l} \phi_{\nu_1\ldots \nu_l} \phi_{\lambda_1\ldots \lambda_l}\phi_{\rho_1\ldots \rho_l} (\eta^{\mu \nu})^{n_{12}}  (\eta^{\mu \lambda})^{n_{13}}  (\eta^{\mu \rho})^{n_{14}}  (\eta^{\nu \lambda})^{n_{23}}  (\eta^{\nu \rho})^{n_{24}}  (\eta^{\lambda \rho})^{n_{34}},
\ee
where $\phi_{\mu_1\ldots \mu_l}$ is the symmetric traceless rank-$l$ tensor field carrying the spin-$l$ particle.  All of the contractions in this vertex are fixed once a choice is made for how the $\mu$ indices are distributed over the other three tensors. Each vertex thus corresponds to a distinct partition of $l$ of length at most three and the number of such partitions is given by Eq.~\eqref{eq:zerodnumber}.

We will focus on the parity-even zero-derivative vertices for higher spins; however, we note that the parity-odd, $S_4$-invariant, zero-derivative vertices are given by the parity-even vertices multiplied by
\be
\left( z_{1 4} z_{2 3} - z_{1 3} z_{2 4} \right) \left(z_{1 4} z_{2 3} - 
   z_{1 2} z_{3 4} \right) \left( z_{1 3} z_{2 4} - z_{1 2} z_{3 4} \right) \varepsilon ( z_1 z_2 z_3 z_4).
\ee
This means that zero-derivative parity-odd vertices exist for spins $\geq4$ and for spin $l$ there are $\left[ (l-1)^2/12 \right]$ of them. 

\section{Four-Point Amplitudes} \label{sec:4particle}

Our goal is to find the most general four-point tree amplitude with a certain energy scaling for some given particle content, and consequently to find a lower bound on the scaling.
Now that we have all the relevant on-shell cubic and quartic vertices from Sec.~\ref{sec:construct}, it is possible to calculate the general 2-to-2 tree amplitude, impose crossing symmetry, and then look for the subset of amplitudes with a certain energy scaling. However, in practice this is quite difficult to implement for particles with spin due to the complexity of the crossing equations and the dependence of the amplitude on multiple free functions.  It is thus helpful to choose simplifying kinematics and to understand in detail how vertices translate into four-point amplitudes. In this section we detail our explicit kinematics and discuss some properties of four-point transversity amplitudes that are needed to organize our calculation. 

\subsection{Kinematics} \label{ssec:kinematics}

We consider identical external particles with mass $m$, so that $p^i\cdot p^i =-m^2$. Particles 3 and 4 are now taken to be outgoing, so momentum conservation gives
\be
p^1+p^2=p^3+p^4.
\ee
In the center-of-mass frame with scattering in the $xz$-plane, scattering angle $\theta$, and center-of-mass energy $E$, the momenta can be written as
\be \label{eq:COMmomenta}
p^j_{ \mu} = \left( E, p \sin \theta_j, 0, p \cos \theta_j \right),
\ee
where $\theta_1=0$, $\theta_2=\pi$, $\theta_3 = \theta$, $\theta_4 = \theta- \pi$. The Mandelstam variables are defined by
\be
s=-(p^1+p^2)^2, \; t=-(p^1-p^3)^2, \; u=-(p^1-p^4)^2,
\ee 
which satisfy $s+t+u=4m^2$. 
These are related to $\theta$ and $E$ by
\be \label{eq:mandeltheta}
s =4 E^2, \quad \cos \theta = 1-\frac{2t}{4m^2 -s} = \frac{2 u}{4m^2 -s}-1, \quad \sin^2 \theta = \frac{4 t u}{(s-4m^2)^2}.
\ee
The limit of high-energy fixed-angle scattering corresponds to $s,-t \rightarrow \infty$ with $s/t$ fixed.

A massive vector has three degrees of freedom. For each momentum $p_{\mu}$, we have a basis of polarization vectors $\epsilon^{(\tau) }_{\mu}$, $\tau=0,\pm 1$, that are transverse and orthonormal, 
\be \epsilon^{(\tau)}_{\mu} p^{\mu} =0,\ \ \epsilon^{(\tau)}_{\mu} \epsilon^{* (\tau') \mu} = \delta^{\tau \tau'}\, .\ee 
We will use a polarization basis corresponding to states with their spins projected on the axis orthogonal to the scattering plane, which are known as transversity states \cite{Kotanski:1965zz}. We could use any polarization basis, but transversity states simplify our calculations due to their simple crossing relations. In the center-of-mass frame \eqref{eq:COMmomenta}, the vector transversity polarizations for external particle $j$ are given by
\begin{subequations} \label{eq:vecpolz}
\begin{align}
\epsilon^{(\pm 1)}_{ \mu}(p^j) & = \frac{i}{\sqrt{2} m} \left( p,E \sin \theta_j \pm i m\cos \theta_j,0,E \cos \theta_j \mp i m \sin \theta_j \right), \\
\epsilon^{(0)}_{ \mu}(p^j) & = \left( 0,0,1,0 \right),
\end{align}
\end{subequations}
where the parenthesized superscript labels the transversity~\cite{deRham:2017zjm}. 
The four external polarization vectors are linear sums of these basis polarizations, 
\be \label{eq:vecpolz1}
\epsilon^j_{\mu}=\sum_{\tau=-1}^1 \alpha_{\tau}^j \epsilon^{(\tau)}_{\mu}(p^j),
\ee
where $ \alpha_{\tau}^j $ are normalized coefficients 
\be \label{eq:vecnorm}
\sum_{\tau=-1}^1 \left|\alpha_{\tau}^j \right|^2 =1.
\ee
The polarization vectors satisfy the completeness relation
\be
\Pi_{\mu \nu} = \sum_{\tau=-1}^1 \epsilon^{(\tau)}_{ \mu} \epsilon_{ \nu}^{(\tau) *},
\ee
where the projection tensor $\Pi_{\mu \nu}$ appears in the numerator of the massive vector propagator and is given by
\be \label{projector}
\Pi_{\mu \nu} = \eta_{\mu \nu} + \frac{p_{\mu} p_{\nu}}{m^2}.
\ee
More generally, the propagator for a massive spin-$l$ particle is given by
\be \label{eq:prop}
\frac{-i \Pi_{\mu_1 \ldots \mu_l, \nu_1 \ldots \nu_l}}{p^2+m^2 - i\epsilon},
\ee
where the numerator is a transverse, traceless projection tensor that is separately symmetric in the $\mu$ and $\nu$ indices.

A massive spin-2 particle has five degrees of freedom. For each momentum $p_{\mu}$, we have a basis of symmetric polarization tensors $\epsilon^{(\tau) }_{\mu \nu}$, $\tau=0,\pm 1, \pm 2$, that are transverse, traceless, and orthonormal, 
\be e^{(\tau)}_{\mu \nu} p^{\mu} =0, \ \ \epsilon^{(\tau) }_{\mu}{}^{\mu}=0, \ \ \epsilon^{(\tau) }_{\mu \nu}\epsilon^{*(\tau') \mu \nu}=\delta^{\tau \tau'}\,.\ee
In terms of the vector polarizations we can take
\begin{subequations}
\begin{align}
\epsilon^{(\pm 2)}_{ \mu \nu} & = \epsilon^{(\pm 1)}_{ \mu} \epsilon^{( \pm 1) }_{\nu}, \\
\epsilon^{(\pm 1)}_{ \mu \nu} & = \frac{1}{\sqrt{2}} \left( \epsilon^{(\pm 1)}_{ \mu} \epsilon^{( 0) }_{\nu}+\epsilon^{(0)}_{ \mu} \epsilon^{( \pm 1) }_{\nu} \right), \\
\epsilon^{(0)}_{ \mu \nu} & = \frac{1}{\sqrt{6}} \left( \epsilon^{(1)}_{ \mu} \epsilon^{( -1) }_{\nu}+ \epsilon^{(-1)}_{ \mu} \epsilon^{( 1) }_{\nu}+2 \epsilon^{(0)}_{ \mu} \epsilon^{( 0) }_{\nu} \right).
\end{align}
\end{subequations}
The four external polarization tensors are linear sums of these basis tensors, 
\be \label{eq:spin2polsum}
\epsilon^j_{\mu \nu}=\sum_{\tau=-2}^2 \alpha_{\tau}^j \epsilon^{(\tau)}_{\mu \nu}(p^j),
\ee
where $ \alpha_{\tau}^j $ are normalized coefficients 
\be
\sum_{\tau=-2}^2 \left|\alpha_{\tau}^j \right|^2 =1.
\ee
The polarization tensors satisfy the completeness relation
\be
\Pi_{\mu_1 \mu_2, \nu_1 \nu_2} = \sum_{\tau=-2}^2 \epsilon^{(\tau)}_{ \mu_1 \mu_2} \epsilon_{ \nu_1 \nu_2}^{(\tau) *},
\ee
where the transverse traceless tensor $\Pi_{\mu_1 \mu_2, \nu_1 \nu_2}$ appears in the numerator of the massive spin-2 propagator and is given by
\be
\Pi_{\mu_1 \mu_2, \nu_1 \nu_2} = \frac{1}{2} \Pi_{\mu_1 \nu_1} \Pi_{\mu_2 \nu_2} + \frac{1}{2} \Pi_{\mu_1  \nu_2} \Pi_{\mu_2 \nu_1} -\frac{1}{3} \Pi_{\mu_1 \mu_2} \Pi_{\nu_1  \nu_2}.
\ee
We can similarly define polarization tensors for spins greater than 2 by building higher-rank symmetrized traceless products of the vector polarizations. 

\subsection{Properties of transversity amplitudes} 
We now review some properties of 2-to-2 transversity amplitudes that will be useful for our calculation. We discuss parity, permutation symmetries, crossing symmetry, kinematical singularities, and the translation of vertices into four-point amplitudes. Our discussion is not general since we assume that the scattered particles are identical bosons with spin $l$. For reviews of transversity amplitudes, see \cite{Kotanski:1970, deRham:2017zjm}.

\subsubsection{Parity}

First consider parity. We denote the four-point amplitude for the scattering of identical particles with transversities $\tau_i$ by $\mathcal{A}_{\tau_1 \tau_2 \tau_3 \tau_4}$. A nice feature of these transversity amplitudes is that they have definite parity, unlike helicity amplitudes. Under a parity transformation they transform as~\cite{deRham:2017zjm}
\be \label{eq:P}
P: \mathcal{A}_{\tau_1 \tau_2 \tau_3 \tau_4} \rightarrow  (-1)^{\tau_1+\tau_2-\tau_3-\tau_4} \mathcal{A}_{\tau_1 \tau_2 \tau_3 \tau_4}.
\ee
So amplitudes with an even or odd sum of transversities are parity even or odd, respectively. The contributions from the parity-odd and parity-even quartic structures thus decouple in the transversity basis. If there are parity-even and parity-odd cubic vertices then the even-even and odd-odd exchange diagrams contribute to the parity-even four-point amplitudes, whereas the even-odd and odd-even exchange diagrams contribute to the parity-odd four-point amplitudes.
In a parity-invariant theory the amplitudes with an odd sum of transversities must vanish.

The number of parity-even and parity-odd amplitudes is
\be \label{eq:totalpolz}
{N}_{\rm total}(l)  =\frac{(2l+1)^4+1}{2}, \quad \tilde{N}_{\rm total}(l) =\frac{(2l+1)^4-1}{2},
\ee
since the number of even transversity sums exceeds the number of odd sums by one.
The number of independent amplitudes will be fewer than this due to the permutation symmetries that we consider next. 

\subsubsection{Permutation symmetries}

We now consider the action of permutation symmetries on transversity amplitudes.  As with the tensor structures, it is useful to consider separately the permutations that leave the Mandelstam variables invariant---the kinematic permutations---and those that do not, which lead to crossing relations. The kinematic permutations act in a simple way on the transversity amplitudes of identical integer-spin particles,
\begin{subequations}
\begin{align} 
(12)(34):&  \quad \mathcal{A}_{\tau_1 \tau_2 \tau_3 \tau_4} =\mathcal{A}_{\tau_2 \tau_1 \tau_4 \tau_3} \label{eq:rotationcon}, \\
(13)(24):& \quad  \mathcal{A}_{\tau_1 \tau_2 \tau_3 \tau_4} = \mathcal{A}_{\tau_3 \tau_4 \tau_1 \tau_2 }, \label{eq:PT} \\
(14)(23):&  \quad \mathcal{A}_{\tau_1 \tau_2 \tau_3 \tau_4} = \mathcal{A}_{\tau_4 \tau_3 \tau_2 \tau_1 }. \label{eq:1423}
\end{align}
\end{subequations}
The first relation follows from the invariance of the amplitude under a rotation by $\pi$ about the axis perpendicular to the center-of-mass scattering plane. To get the second relation, note that invariance under time reversal $T$ gives~\cite{deRham:2017zjm}
\be \label{eq:T}
T \text{ invariance} \implies \mathcal{A}_{\tau_1 \tau_2 \tau_3 \tau_4}  =  (-1)^{\tau_1-\tau_2-\tau_3+\tau_4} \mathcal{A}_{\tau_3 \tau_4 \tau_1 \tau_2}\,.
\ee
Time reversal is not a symmetry of our parity-odd amplitudes, but charge conjugation $C$ is a symmetry since our particles are identical and uncharged. By the $CPT$ theorem we must therefore have $PT$ as a symmetry, which from Eqs.~\eqref{eq:P} and \eqref{eq:T} implies Eq.~\eqref{eq:PT}. The third relation, Eq. \eqref{eq:1423}, follows by combining the other two. 

These permutation relations reduce the number of independent amplitudes. We can count the number of independent parity-even and parity-odd amplitudes using Burnside's lemma, which gives
\begin{align}
N_{\rm amp.}(l)& = 1+2 l(1+l)(2+l+l^2),\\
\tilde{N}_{\rm amp.}(l) &= l(1+l)(1+2l+2l^2) \label{eq:Ncalodd}.
\end{align}
Notice that the number of independent amplitudes precisely matches the number of independent quartic structures from Eqs. \eqref{eq:Neven} and \eqref{eq:Nodd},
\be
N_{\rm amp.}(l) =N_\text{quartic}(l), \quad \tilde{N}_{\rm amp.}(l)=\tilde{N}_\text{quartic}(l).
\ee
This equivalence is apparent from the explicit construction of tensor structures in \cite{Kravchuk:2016qvl} and is explored more generally in App.~F of~\cite{Henning:2017fpj}. The implication of this is that there is one functional degree of freedom in the amplitude per independent scattering process.  Placing an upper bound on the growth of each process is thus constraining enough to determine the functions up to a finite number of parameters. 

\subsubsection{Crossing symmetry}

Now we consider the four-particle crossing symmetries, which correspond to the permutation symmetries that change the Mandelstam variables. The primary advantage of transversity amplitudes is their simple crossing symmetry transformations. For the scattering of identical bosons the crossing relations are \cite{Kotanski:1965zz, deRham:2017zjm}
\begin{align}
\mathcal{A}_{\tau_1 \tau_2 \tau_3 \tau_4}(s,t) & = e^{i\left(\pi- \chi_t\right) \sum_j \tau_j }\mathcal{A}_{-\tau_1 -\tau_3 -\tau_2 -\tau_4}(t,s), \label{eq:stcrossing}\\
\mathcal{A}_{\tau_1 \tau_2 \tau_3 \tau_4}(s,t) & = e^{i \left( \pi- \chi_u\right) \sum_j \tau_j }\mathcal{A}_{-\tau_1 -\tau_4 -\tau_3 -\tau_2}(u,t), \label{eq:sucrossing}
\end{align}
where
\begin{align}
e^{-i \chi_t} = \frac{-st -2 i m \sqrt{s t u}}{\sqrt{ s(s-4m^2)t(t-4m^2)}}, \quad e^{-i \chi_u} = \frac{-su +2 i m \sqrt{s t u}}{\sqrt{ s(s-4m^2)u(u-4m^2)}}.
\end{align}
The crossing relations \eqref{eq:stcrossing} and \eqref{eq:sucrossing} require analytically continuing the amplitudes, since the left- and right-hand sides correspond to physical scattering in different regions of the complex Mandelstam plane. In practice we evaluate everything in the physical $s$-channel region where $s=s_0+i \epsilon$,  $t=t_0-i \epsilon$, and $u<0$ with $s_0>4m^2$, $t_0<0$, and $\epsilon \rightarrow 0^+$. We have checked that Eqs. \eqref{eq:stcrossing} and \eqref{eq:sucrossing} hold for a few test amplitudes for self-interacting spin-1 and spin-2 particles derived from Lagrangian interactions. For a recent derivation and discussion of the crossing relations for spinning particles, see \cite{deRham:2017zjm}.

\subsubsection{Singularity structure}

Amplitudes for spinning particles can contain singularities beyond the usual poles and branch cuts of scalar amplitudes. The structure of these kinematical singularities depends on the relative masses of the external particles in an intricate way~\cite{Cohen-Tannoudji:1968lnm,Kotanski1968}.  In the case of identical integer-spin particles we can extract the kinematical singularities by writing the transversity amplitudes in the form
\be \label{eq:ampexpand}
\mathcal{A}_{\tau_1 \tau_2 \tau_3 \tau_4}(s,t) = \frac{s^{\xi} a_{\tau_1 \tau_2 \tau_3 \tau_4}(s, t)+i \sqrt{s t u} \, b_{\tau_1 \tau_2 \tau_3 \tau_4}( s,t)}{\sqrt{s}^{\; \xi}\left(s-4m^2\right)^{ |\sum_i \tau_i|/2}},
\ee
where $\xi=0$ if the amplitude is parity even and $\xi=1$ if the amplitude is parity odd. The $\sqrt{s-4m^2}$, $\sqrt{s}$, and $\sqrt{s t u}$ singularities are called thresholds, pseudothresholds, and ``the border of the physical region," respectively. The threshold and pseudothreshold singularities factorize for transversity amplitudes, as can be deduced from the crossing relations, whereas the $\sqrt{s t u}$ piece does not factorize~\cite{Cohen-Tannoudji:1968lnm,Kotanski1968,Kotanski:1970}. 

The functions $a_{\tau_1 \tau_2 \tau_3 \tau_4}(s, t)$ and $b_{\tau_1 \tau_2 \tau_3 \tau_4}(s, t)$, which are defined by Eq.~\eqref{eq:ampexpand}, encode the amplitudes with the kinematical singularities stripped off. They are not all independent due to kinematic permutations and the following additional relation under flipping transversities:
\be \label{eq:tranflip}
\mathcal{A}_{-\tau_1 -\tau_2 -\tau_3 -\tau_4}(s,t) = \frac{(-1)^{\xi} \left(s^{\xi} a_{\tau_1 \tau_2 \tau_3 \tau_4}(s, t)-i \sqrt{s t u} \, b_{\tau_1 \tau_2 \tau_3 \tau_4}( s,t)\right)}{\sqrt{s}^{\; \xi}\left(s-4m^2\right)^{ |\sum_i \tau_i|/2}}.
\ee
This implies, for example, that $b_{0000}(s,t)=0$. We can use Burnside's lemma to count the number of independent parity-even and parity-odd functions $a_{\tau_1 \tau_2 \tau_3 \tau_4}(s, t)$ and $b_{\tau_1 \tau_2 \tau_3 \tau_4}(s, t)$,
\begin{align}
N_{a}(l) = \frac{2+l(1+l)(7+2l+2l^2)}{2}, \quad \tilde{N}_{a}(l) =N_{b}(l)=\tilde{N}_{b}(l)= \frac{\tilde{N}_{\rm amp.}(l)}{2}.
\end{align}
The total number of these functions is equal to the number of independent amplitudes, 
\be
N_{a}(l)+N_{b}(l) =N_{\rm amp.}(l) , \quad \tilde{N}_{a}(l)+\tilde{N}_{b}(l)=\tilde{N}_{\rm amp.}(l) ,
\ee
so they correctly account for all of the functional degrees of freedom.

\subsubsection{Translating vertices into amplitudes}

To relate on-shell cubic and quartic vertices to these four-point amplitudes, we can split the functions $a_{\tau_1 \tau_2 \tau_3 \tau_4}(s,t)$ into pieces coming from tree-level exchange and contact diagrams. We thus write
\be \label{eq:asplit}
a_{\tau_1 \tau_2 \tau_3 \tau_4}(s,t) = a^{\rm exchange}_{\tau_1 \tau_2 \tau_3 \tau_4}(s,t)+a^{\rm contact}_{\tau_1 \tau_2 \tau_3 \tau_4}(s,t),
\ee
where $a^{\rm exchange}_{\tau_1 \tau_2 \tau_3 \tau_4}(s,t)$ is a rational function of $s$ and $t$ that has simple poles corresponding to exchanged states going on shell and depends quadratically on the cubic couplings $g_m$, $\tilde{g}_{\tilde{m}}$, as fixed by factorization. 
Similarly, $a^{\rm contact}_{\tau_1 \tau_2 \tau_3 \tau_4}(s,t)$ is a polynomial that depends linearly on the tensor structure coefficients $f_I(s,t)$, $\tilde{f}_{\tilde{I}}(s,t)$ defined in Eqs.~\eqref{eq:genquartic} and \eqref{eq:genquarticodd},
\be \label{eq:a4expansion}
   a^{\rm contact}_{\tau_1 \tau_2 \tau_3 \tau_4}(s,t) = 
\begin{cases}
   \sum_{I} q^{I}_{\tau_1 \tau_2 \tau_3 \tau_4}(s,t) f_I(s,t) &  \text{for } \sum \tau \text{ even,} \\
     \sum_{\tilde{I}} \tilde{q}^{\tilde{I}}_{\tau_1 \tau_2 \tau_3 \tau_4}(s,t) \tilde{f}_{\tilde{I}}(s,t) & \text{for } \sum \tau \text{ odd,}
\end{cases}
\ee
where $q^{I}_{\tau_1 \tau_2 \tau_3 \tau_4}(s,t)$, $\tilde{q}^{\tilde{I}}_{\tau_1 \tau_2 \tau_3 \tau_4}(s,t)$ are polynomials that encode the quartic tensor structures. 

We similarly write
\be
b_{\tau_1 \tau_2 \tau_3 \tau_4}(s,t)  = b^{\rm exchange}_{\tau_1 \tau_2 \tau_3 \tau_4}(s,t)+b^{\rm contact}_{\tau_1 \tau_2 \tau_3 \tau_4}(s,t), 
 \label{eq:b3expansion}
\ee
and
\be
 \label{eq:b4expansion}
   b^{\rm contact}_{\tau_1 \tau_2 \tau_3 \tau_4}(s,t)  = 
\begin{cases}
   \sum_{I} r^{I}_{\tau_1 \tau_2 \tau_3 \tau_4}(s,t) f_I(s,t) &  \text{for } \sum \tau \text{ even}, \\
     \sum_{\tilde{I}} \tilde{r}^{\tilde{I}}_{\tau_1 \tau_2 \tau_3 \tau_4}(s,t) \tilde{f}_{\tilde{I}}(s,t)   & \text{for } \sum \tau \text{ odd,}
\end{cases}
\ee
where $r^{I}_{\tau_1 \tau_2 \tau_3 \tau_4}(s,t)$, $\tilde{r}^{\tilde{I}}_{\tau_1 \tau_2 \tau_3 \tau_4}(s,t) $ are polynomials that encode the quartic tensor structures. 

\section{Lower Bounds on the Growth of Tree Amplitudes} 
\label{sec:method+results}

With the setup from the previous sections, we can now describe a procedure for finding a lower bound on the growth of tree amplitudes. We first explain our method in detail and then present the results of applying this to theories with a single massive particle.

\subsection{Details of our method}
\label{ssec:detailsofmethod}

Given all the relevant cubic and quartic vertices, an algorithm for finding the general four-point tree amplitude with an arbitrary but finite number of derivatives that grows with energy as $E^n$ for integer $n$ is the following:
\begin{enumerate}
\item If there are on-shell cubic vertices, calculate the contribution to the four-point amplitude from tree-level exchange diagrams using the general cubic vertex. This determines the functions $a^{\rm exchange}_{\tau_1 \tau_2 \tau_3 \tau_4}(s,t)$ and $b^{\rm exchange}_{\tau_1 \tau_2 \tau_3 \tau_4}(s,t)$ in terms of the cubic couplings $g_m$, $\tilde{g}_{\tilde{m}}$. 
\item Write down an ansatz for $a^{\rm contact}_{\tau_1 \tau_2 \tau_3 \tau_4}(s,t)$ as a polynomial in $s$ and $t$ of order 
\be
j_{\rm max} \equiv \floor*{\frac{(n-\xi+\left|\sum_{i=1}^4 \tau_i \right|)}{2}}
\ee
with arbitrary constant coefficients $\alpha^{k}_{\tau_1 \tau_2 \tau_3 \tau_4}$,
\be \label{eq:a4cubicseries}
a^{\rm contact}_{\tau_1 \tau_2 \tau_3 \tau_4}(s,t) = \sum_{j=0}^{j_{\rm max}}\sum_{i=0}^j \alpha^{i+\frac{j(j+1)}{2}}_{\tau_1 \tau_2 \tau_3 \tau_4} s^i t^{j-i}.
\ee
\item Take $a^{\rm exchange}_{\tau_1 \tau_2 \tau_3 \tau_4}(s,t)$ from step 1 and, if possible, Taylor expand around $s, -t=\infty$ with $s/t$ fixed down to order $j_{\rm max}+1$, 
\be \label{eq:a3cubicseries}
a^{\rm exchange}_{\tau_1 \tau_2 \tau_3 \tau_4}(s,t)\Big|_{\text{high energy}} = \sum_{j=j_{\rm max}+1}^{j'_{\rm max}}\sum_{i=0}^j \overline{\alpha}^{i+\frac{j(j+1)}{2}}_{\tau_1 \tau_2 \tau_3 \tau_4} s^i t^{j-i} + \text{lower order terms},
\ee
where $j'_{\rm max}$ is the maximum order that appears from the finite number of cubic vertices and $\overline{\alpha}^{k}_{\tau_1 \tau_2 \tau_3 \tau_4}$ are coefficients that depend only on the cubic couplings $g_m$, $\tilde{g}_{\tilde{m}}$.\footnote{Writing the exchange amplitude as in \eqref{eq:a3cubicseries} may not be possible for a given spectrum of particles since the expansion eventually has nonpolynomial pieces. This means that the exchange terms cannot be cancelled by contact terms and we must add additional particles to achieve the desired tree-level high-energy behavior.} 
\item Update the ansatz for $a^{\rm contact}_{\tau_1 \tau_2 \tau_3 \tau_4}(s,t)$ by subtracting from \eqref{eq:a4cubicseries} the first term on the right-hand side of Eq. \eqref{eq:a3cubicseries},
\be
a^{\rm contact}_{\tau_1 \tau_2 \tau_3 \tau_4}(s,t) = \sum_{j=0}^{j_{\rm max}}\sum_{i=0}^j \alpha^{i+\frac{j(j+1)}{2}}_{\tau_1 \tau_2 \tau_3 \tau_4} s^i t^{j-i} - \sum_{j=j_{\rm max}+1}^{j'_{\rm max}}\sum_{i=0}^j \overline{\alpha}^{i+\frac{j(j+1)}{2}}_{\tau_1 \tau_2 \tau_3 \tau_4} s^i t^{j-i}.
\ee
This ensures that $a_{\tau_1 \tau_2 \tau_3 \tau_4}(s,t)$ in Eq. \eqref{eq:asplit} is bounded by order $j_{\rm max}$ in $s$ and $t$ at high energy, since the contact terms now cancel off the high-energy behavior of the exchange terms.
\item Repeat the previous three steps with the replacements $a \rightarrow b$, $n \rightarrow n-3$, and $\xi\rightarrow -\xi$. From Eq.~\eqref{eq:ampexpand}, this produces an ansatz for the total amplitude $\mathcal{A}_{\tau_1 \tau_2 \tau_3 \tau_4}(s,t)$ with the desired $E^n$ behavior.
\item Next impose crossing symmetry. Substitute the ansatz for $\mathcal{A}_{\tau_1 \tau_2 \tau_3 \tau_4}(s,t)$ into the crossing equations \eqref{eq:stcrossing} and \eqref{eq:sucrossing}. By separating the parts that are and are not proportional to $\sqrt{s t u}$, the crossing relations can be written as polynomial equations in $s$, $t$ with coefficients that depend linearly on $a^{k}_{\tau_1 \tau_2 \tau_3 \tau_4}$, $b^{k}_{\tau_1 \tau_2 \tau_3 \tau_4}$ and quadratically on the cubic couplings $g_m$, $\tilde{g}_{\tilde{m}}$. Solve these constraints at each order in $s$ and $t$ and update the ansatz accordingly. It is helpful to transform the constraints into a linear system by replacing products of cubic couplings with new variables, e.g. $g_m g_n \rightarrow g_{m n}$, and then to impose the constraint $g_{m n} = g_m g_n$ at the end of step 8.
\item Now make use of the explicit form of the Lorentz-invariant quartic vertices. Calculate the four-point amplitude from the general contact vertices \eqref{eq:genquartic} and \eqref{eq:genquarticodd} to determine the polynomials $q^{I}_{\tau_1 \tau_2 \tau_3 \tau_4}(s,t)$, $\tilde{q}^{\tilde{I}}_{\tau_1 \tau_2 \tau_3 \tau_4}(s,t)$ and $r^{I}_{\tau_1 \tau_2 \tau_3 \tau_4}(s,t)$, $\tilde{r}^{\tilde{I}}_{\tau_1 \tau_2 \tau_3 \tau_4}(s,t)$ in the expansions \eqref{eq:a4expansion} and \eqref{eq:b4expansion}.
\item Finally, enforce consistency of the contact ansatz with the explicit form of the quartic vertices. For example, in the parity-even case the ansatz for $a^{\rm contact}_{\tau_1 \tau_2 \tau_3 \tau_4}(s,t)$ must satisfy the consistency condition
\be \label{eq:consistency}
a^{\rm contact}_{\tau_1 \tau_2 \tau_3 \tau_4}(s,t)=    \sum_{I} q^{I}_{\tau_1 \tau_2 \tau_3 \tau_4}(s,t) \sum_{j=0}^{\infty }\sum_{i=0}^j f_I^{i+\frac{j(j+1)}{2}} s^i t^{j-i},
\ee 
where we have Taylor expanded $f_I(s,t)$. At each order in $s$ and $t$ this gives a consistency condition involving the Taylor series coefficients $f_I^k$ and the remaining quartic coefficients $\alpha^{k}_{\tau_1 \tau_2 \tau_3 \tau_4}$ and cubic couplings $g_m$, $\tilde{g}_{\tilde{m}}$. There are similar conditions in the parity-odd case and for $b^{\rm contact}_{\tau_1 \tau_2 \tau_3 \tau_4}(s,t)$. The last step is to simultaneously solve these consistency conditions and update the ansatz.\footnote{There are an infinite number of these consistency conditions if the set of tensor structures is not a basis since then the functions $f_I(s,t)$, $\tilde{f}_{\tilde{I}}(s,t)$ can be unbounded, even though the sum in Eq. \eqref{eq:consistency} is bounded. In practice we then only solve the finite number of constraints involving $a^{k}_{\tau_1 \tau_2 \tau_3 \tau_4}$, $b^{k}_{\tau_1 \tau_2 \tau_3 \tau_4}$ and $g_m$, $\tilde{g}_{\tilde{m}}$. This gives necessary conditions on the amplitude and we know that these conditions are sufficient if we can find a theory realizing the amplitude. }
\end{enumerate}
Throughout we also require that all of the cubic couplings and quartic coefficients are real (once the usual factors of $i$ are inserted in the vertices), as required by unitarity.
This algorithm then produces the most general four-point tree-level amplitude with the required scaling at high energy that is consistent with Lorentz invariance, locality, unitarity, and crossing symmetry.\footnote{There are additional restrictions on EFT amplitudes that we do not impose here, such as positivity constraints and the absence of a Shapiro time advance \cite{Adams:2006sv, Camanho:2014apa, Bellazzini:2016xrt, Cheung:2016yqr, Bonifacio:2016wcb, deRham:2017zjm, Hinterbichler:2017qyt,Bonifacio:2017nnt,Hinterbichler:2017qcl}. Note also that improved positivity bounds can imply that the EFT cutoff is much lower than what would naively be inferred from the strong coupling scale~\cite{Bellazzini:2017fep, deRham:2017xox}.} 

\subsection{Results}
In this section we present the results of applying the above procedure to a theory containing a single massive particle with spin 0, 1, or 2. We also give a conjecture for a lower bound on the growth of four-point amplitudes for a single massive higher-spin particle. 

\subsubsection{Spin 0}
We begin with the simple case of four-point scalar scattering. Although the general algorithm is excessive in this case and the result is well known, we apply it to illustrate the procedure. There are no kinematical singularities and the general amplitude is just
\be
\mathcal{A}_{0000}(s,t) =  a_{0000}(s, t) =  a^{\rm exchange}_{0000}(s,t)+a^{\rm contact}_{00 00}(s,t).
\ee
The constant cubic vertex $\lambda$ contributes to $a^{\rm exchange}_{0000}(s,t)$ through $s$-, $t$-, and $u$-channel exchange diagrams,
\be
a^{\rm exchange}_{0000}(s,t) = -\lambda^2 \left(\frac{1}{s-m^2}+\frac{1}{t-m^2} + \frac{1}{u-m^2} \right).
\ee
This does not grow with energy, so there is nothing that needs cancelling by contact terms. 

There is a single constant quartic structure with coefficient $f_1(s,t) =a^{\rm contact}_{00 00}(s,t)$. Crossing symmetry requires that $f_1(s,t)$ is invariant under interchanging $s$, $t$ and $u$. By the fundamental theorem of symmetric polynomials, we can write the general polynomial solution as
\be \label{eq:scalarquartic}
a^{\rm contact}_{00 00}(s,t) = \sum_{\substack{i,j\geq0 \\ 2i+3j \leq n/2}} \alpha_{ij} (st+su+tu)^i (stu)^j,
\ee
where $\alpha_{ij}$ are constants and $E^n$ is the desired energy scaling. The best UV behavior is produced by an amplitude with couplings $\lambda$ and $\alpha_{00}$, corresponding to the renormalizable theory with $\phi^3$ and $\phi^4$ couplings.

\subsubsection{Spin 1}

Next we consider amplitudes for a massive spin-1 particle. In this case there are multiple tensor structures at quartic order but without the added complication of cubic vertices. Since there are no exchange diagrams, the contributions from parity-odd and parity-even vertices are decoupled in the transversity basis. 

Consider a massive vector theory with a quartic potential, given by the Lagrangian
\be
\mathcal{L} = -\frac{1}{4} F_{\mu \nu} F^{\mu \nu} -\frac{1}{2} m^2 A^2+ \lambda_1  (A^2)^2,
\ee
where $A^2 \equiv A_{\mu} A^{\mu}$. The amplitudes in this theory grow at most as $E^4$ (for scattering longitudinally polarized modes), so we are mostly interested in theories with amplitudes that grow at least this slowly.

Applying the algorithm described in Sec.~\ref{ssec:detailsofmethod} using the general quartic amplitude \eqref{eq:spin1quartic}, we find that there is a seven-parameter family of amplitudes that have $E^4$ scaling and that improving this is impossible without additional particles.\footnote{With multiple massive spin-1 particles it is possible to do better; e.g. non-Abelian Yang-Mills theory with a mass term has amplitudes that grow at worst like $E^2$ \cite{ArkaniHamed:2002sp}.
} Five of the free parameters are parity even, $\lambda_k$, $k=1,\ldots, 5$, and two are parity odd, $\tilde{\lambda}_{\tilde{k}}$, $\tilde{k}=1,2$.
Since we have a basis of quartic structures, we can characterize the solution by giving the coefficients of the tensor structures, $f_I(s,t)$ and $\tilde{f}_{\tilde{I}}(s,t)$. We list these explicitly in Appendix~\ref{app:solutions}. We can also summarize the result by writing down a Lagrangian that generates the amplitude,
\begin{align}
 \mathcal{L} & =  \lambda_1 ( A^2 )^2 + \lambda_2 A^2{F}_{\mu \nu} F^{\mu \nu}+\lambda_3 A_{\mu}A_{ \nu}\left( \partial^{\mu} A_{\lambda} \partial^{\nu} A^{\lambda}- \partial_{\lambda} A^{\mu} \partial^{\lambda} A^{\nu}\right) \nonumber \\ & +\lambda_4 \left( {F}_{\mu \nu} F^{\mu \nu} \right)^2 + \lambda_5 F_{\mu}{}^{\nu}F_{\nu}{}^{\rho} F_{\rho}{}^{\lambda}F_{\lambda}{}^{\mu} 
+ \tilde{\lambda}_1 A^2 \widetilde{F}_{\mu \nu} F^{\mu \nu} + \tilde{\lambda}_2 \widetilde{F}_{\mu \nu} F^{\mu \nu} F_{\alpha \beta} F^{\alpha \beta} \label{eq:veclambda2},
\end{align} 
where $\tilde{F}^{\mu \nu} \equiv \varepsilon^{\mu \nu \alpha \beta} F_{\alpha \beta}$. 
In addition to the quartic potential, there are two parity-even two-derivative terms, the two four-derivative Euler-Heisenberg terms, and two parity-odd terms. The three four-derivative terms represent all quartic gauge-invariant terms. These terms represent precisely the quartic-order ghost-free terms from generalized Proca theories~\cite{Tasinato:2014eka,Heisenberg:2014rta,Jimenez:2016isa}, so ghost-freeness coincides with having tree amplitudes that saturate the lower bound on the growth up this order.\footnote{We thank Lavinia Heisenberg for pointing this out.}

We can also generalize this result by studying how the number of free parameters changes as we alter $n$ in the scaling $E^n$. The numbers of parity-even and parity-odd interactions with energy scaling at most $E^n$ for different $n$ are given in the following table:
\begin{center}
  \begin{tabular}{ l | c c c c c c }
    $n$ & 3 & 4 & 5& 6 & 7 & 8\\ \hline
    Parity-even interactions & 0 & 5& 5& 12 & 13 & 21 \\ 
    Parity-odd interactions & 0 & 2 & 2& 6 & 7 & 10 \\
  \end{tabular}
\end{center}

\subsubsection{Spin 2}

Now we consider massive spin-2 scattering amplitudes. This is the most computationally difficult case that we consider due to the existence of cubic vertices and a large number of quartic vertices. There are two known ghost-free theories of a massive spin-2 particle, dRGT massive gravity \cite{deRham:2010ik, deRham:2010kj} and the pseudolinear theory \cite{Folkerts:2011ev, Hinterbichler:2013eza}. These can be thought of as massive spin-2 theories where the mass term either breaks full or linearized diffeomorphism symmetry. The dRGT theory consists of the Einstein-Hilbert Lagrangian plus special zero-derivative potential terms, which can be parameterized by two couplings conventionally called $c_3$ and $d_5$ (for reviews of massive gravity, see~\cite{Hinterbichler:2011tt, deRham:2014zqa}). The kinetic term in the pseudolinear theory is the linear Fierz-Pauli Lagrangian \cite{Fierz:1939ix} and the interactions are
\be \label{eq:pseudolinear}
\mathcal{L}_{\rm int.}^{\rm pseudolinear} =\frac{1}{M_p}\varepsilon^{\mu_1 \mu_2 \mu_3 \mu_4} \varepsilon^{\nu_1 \nu_2 \nu_3 \nu_4} \left( \lambda_1 \partial_{\mu_1}\partial_{\nu_1} + m^2 \lambda_2 \eta_{\mu_1 \nu_1} + \frac{m^2}{M_p}\lambda_3  h_{\mu_1 \nu_1}\right) h_{\mu_2 \nu_2} h_{\mu_3 \nu_3} h_{\mu_4 \nu_4}  ,
\ee
where $\lambda_1,\lambda_2,\lambda_3$ are coupling constants~\cite{Hinterbichler:2013eza}.
 In both cases the $2 \rightarrow 2$ amplitudes\footnote{The four-point dRGT amplitude can be found in the ancillary file to Ref. \cite{Cheung:2016yqr}.} grow as $E^6$ at high energy and become strongly coupled at the scale $\Lambda_3 = \left(m^2 M_p \right)^{1/3}$. We are thus most interested in looking for amplitudes that grow at least as slowly as $E^6$.  

Going through the general algorithm, we must include the following ingredients: four parity-even cubic vertices, two parity-odd cubic vertices, $201$ parity-even quartic vertices, and 462 parity-odd quartic vertices. After a lengthy calculation, we find two separate amplitudes with $E^6$ scaling. In these amplitudes the cubic couplings are constrained to satisfy either
\be
g_3 = 2 g_2, \quad g_5=\tilde{g}_1=\tilde{g}_5=0,
\ee 
or
\be
g_3 =  g_2, \quad g_5=\tilde{g}_1=\tilde{g}_5=0,
\ee
and each amplitude has three free parameters.
Moreover, all of the parity-odd amplitudes must vanish,
\be
\mathcal{A}_{\tau_1 \tau_2 \tau_3 \tau_4}(s,t)=0 \quad \text{for } \sum_i \tau_i \text{ odd},
\ee
so parity conservation follows automatically from the $E^6$ requirement. 

A comparison shows that the amplitudes correspond precisely to dRGT massive gravity (which has the two parameters $c_3$ and $d_5$ plus the freedom to rescale the Planck mass) and the pseudolinear theory, with no additional interactions. Since there is no choice of parameters in these theories for which the amplitudes grow more slowly, this is also the best possible behavior for a theory with a single massive spin-2 particle. This suggests a close connection between good inferred high-energy behavior and the absence of ghosts. It also implies that no additional interactions can be added to the ghost-free theories without making the four-point amplitudes more divergent at high energy. 

We can also look for spin-2 amplitudes that are slightly more divergent at high energies. At order $E^7$ we find that there are two families of amplitudes with either parity-even cubic interactions (six free parameters) or parity-odd cubic interactions (four free parameters). The amplitudes with parity-even cubics have $g_5 =0$ but $g_1$, $g_2$ and $g_3$ can take any relative value, so this becomes either dRGT or the pseudolinear theory upon further restricting to $E^6$.  The case with parity-odd cubic couplings has $\tilde{g_1} \neq 0$ and $\tilde{g}_5 =0$. Each case also permits a single parity-violating quartic interaction with two derivatives. Looking for the corresponding Lagrangian interaction, we find that it is given by
\be \label{eq:parityoddquartic}
\mathcal{L}_{\text{parity-violating}}=\tilde{\lambda}_1 \epsilon^{\mu_1 \mu_2 \mu_3 \mu_4} h_{\mu_1 \nu_1} h_{\mu_2 \nu_2} \left( \partial^{\nu_3} h_{\mu_3}{}^{\nu_1} \partial_{\nu_3} h_{\mu_4}{}^{\nu_2}-\partial^{\nu_1} h_{\mu_3 \nu_3} \partial^{\nu_2} h_{\mu_4}{}^{\nu_3}\right).
\ee
It would be interesting to look for a ghost-free spin-2 theory that makes use of the parity-odd vertices in \eqref{eq:parityoddgraviton} and \eqref{eq:parityoddquartic}.

\subsubsection{Spin $l$}
Lastly, we look for a lower bound on the growth of zero-derivative quartic amplitudes for general integer spins. Since we use $S_4$-invariant quartic vertices, the crossing relations are automatically satisfied. This means that we can just calculate the quartic amplitude using any polarization basis and fix the couplings so that the amplitude has the slowest growth. 

For spin-1 the only zero-derivative vertex is given by
\be
\mathcal{M}_1 = z_{12}z_{34}+z_{13}z_{24} +z_{14}z_{23}.
\ee
This corresponds to the $(A^2)^2$ term in \eqref{eq:veclambda2} and scales as $E^4$. For spin-2 there are two zero-derivative quartic structures and the amplitude that grows most slowly is
\begin{align}
\mathcal{M}_2&  =  2 \left( z_{12} z_{13} z_{24} z_{34}+ z_{12} z_{14} z_{23} z_{24} + z_{13} z_{14} z_{23} z_{24} \right) - \left(  z_{12}^2 z_{34}^2 +  z_{13}^2 z_{24}^2 +  z_{14}^2 z_{23}^2 \right),\\
& = \varepsilon \left(z_1 z_2 z_3 z_4 \right)\varepsilon \left(z_1 z_2 z_3 z_4 \right).
\end{align}
This corresponds to the quartic pseudolinear interaction from \eqref{eq:pseudolinear} and scales as $E^6$. 

We conjecture that for spin $l$ the zero-derivative quartic amplitudes with the slowest growth are given by products of the low-spin amplitudes,
\be
\mathcal{M}_{l} = \begin{cases} \mathcal{M}_2^{l/2} &  \text{for } l \ {\rm even,}  \\  \mathcal{M}_1 \mathcal{M}_2^{(l-1)/2} &  \text{for } l\  {\rm odd,}\end{cases}
\ee
and that these scale with energy like $E^{3 l}$  or $E^{3l+1}$ for even or odd spin, respectively.
We have explicitly checked that this is true for $l \leq 4$. We further speculate that this growth is the slowest possible, even allowing for derivative interactions. This is reasonable because derivatives tend to increase the growth and this gives the correct general result for $l \leq 2$. 

\section{Discussion}
\label{sec:discussion}

We have presented a method for directly constructing tree-level scattering amplitudes for massive particles with spin, without recourse to a Lagrangian, by enforcing the requirements of Lorentz invariance, locality, unitarity, and crossing symmetry.  We used this method to find model-independent lower bounds on the growth of tree-level amplitudes in effective field theories containing a single massive particle with integer spin, allowing for all possible self-interactions containing an arbitrary but finite number of derivatives. We proved a general bound for spins $\leq 2$ and conjectured a bound for spins $>2$.  Although we worked in four dimensions, the method we used works in arbitrary dimensions. Our calculation in four dimensions could likely be simplified by using the massive spinor-helicity formalism discussed in Ref.~\cite{Arkani-Hamed:2017jhn}.

Considering an EFT of a single massive spin-2 particle, we have shown that $E^6$ is a lower bound on the growth with energy of the four-point amplitude. In a scheme where fields are scaled with $M_P$ and derivatives with the mass $m$, this corresponds to a highest possible strong coupling scale of $\Lambda_3 = (M_Pm^2)^{1/3}$.  This bound is saturated only by dRGT massive gravity and the spin-2 pseudolinear theory. This suggests that saturating the lower bound on energy growth is a feature that characterizes dRGT scattering amplitudes, which is analogous to how enhanced soft limits characterize amplitudes for certain massless scalar EFTs~\cite{Cheung:2014dqa}. It would be interesting to explore whether or not this holds for higher-point amplitudes.

When looking for effective field theories for use in cosmology, there is often an emphasis on nonperturbative ghost freedom, meaning that the fully nonlinear theory should propagate the same number of degrees of freedom as the linear theory around some standard background.  From the point of view of the perturbative $S$-matrix, the notion of ghost freedom does not seem to be an intrinsic property, because order-by-order field redefinitions that leave the S-matrix invariant \cite{Haag:1958vt} can make a ghost-free theory look ghostly and vice versa \cite{Burgess:2014lwa}, and it is not always possible to field redefine in this way to get a ghost-free structure \cite{Solomon:2017nlh}.  However, in the context of interacting massive theories with spin, nonlinearly ghost-free 
Lagrangians appear to be associated with $S$-matrices that saturate the lower bound on the amplitude growth. One example is dRGT massive gravity and the pseudolinear theory: these are both nonlinearly ghost free \cite{Hassan:2011hr,Hassan:2011ea,Mirbabayi:2011aa,Hinterbichler:2013eza} and, as we have seen, they generate amplitudes with the lowest possible energy scaling for a self-interacting massive spin-2 particle. Another example is the Federbush Lagrangian describing a massive spin-2 particle interacting with an Abelian gauge field~\cite{Federbush1961}, which is also nonlinearly ghost free \cite{deRham:2014tga} and saturates the lower bound for the energy growth of tree amplitudes among these degrees of freedom \cite{Porrati:2008an,Porrati:2008ha}.  We expect that similar statements apply to ghost-free bigravity \cite{Hassan:2011zd} and multigravity \cite{Hinterbichler:2012cn}.  It would be useful to find a deeper explanation for this connection and to know if it extends to higher-spin degrees of freedom.

We focused in this paper on theories that contain a single massive particle, but our methods can easily be generalized to include more complicated particle spectra. One example would be to include cubic couplings of a massive spin-2 particle to additional low-spin states, as investigated for the case of relevant and marginal operators in~\cite{Christensen:2014wra}. Such states then contribute to massive spin-2 scattering through exchange diagrams and could help cancel the high-energy behavior of amplitudes beyond what we have found. This is the mechanism by which the scalar Higgs field prevents high-energy violation of perturbative unitarity in massive spin-1 amplitudes. 
For a massive spin-2 particle there is no known Higgs theory with a finite number of particles and bounded high-energy behavior and there are arguments against the existence of such a theory~\cite{Arkani-Hamed:2017jhn}, although in Kaluza-Klein theories the massive spin-2 amplitudes are partially improved~\cite{Schwartz:2003vj}. 
It may be necessary to include infinitely many high spin intermediate states to obtain amplitudes that decay in the UV, as in large-$N$ QCD-like theories and (if we include gravity) string theory. It is an open problem whether tree-level string amplitudes are the \textit{only} weakly coupled completions of gravity amplitudes; for recent work see, e.g. Refs.~\cite{Caron-Huot:2016icg, Sever:2017ylk, Nayak:2017qru}. Making further progress on this will require better understanding general scattering amplitudes including massive external states with spin, so some of the methods used in this paper could be helpful for such calculations.

\vspace{-.4cm}
\paragraph{Acknowledgements:} For helpful discussions and feedback we would like to thank Clifford Cheung, Francesca Day, George Johnson, Austin Joyce, Scott Melville, Rachel Rosen, James Scargill, Andrew Tolley, and especially Claudia de Rham.

\appendix

\section{Spin-1 Quartic Vertices}
\label{app:structures}

In this appendix we give explicit bases for the spin-1 quartic tensor structures with all momenta incoming. 
A basis of the 17 $\mathbb{Z}_2^2$-invariant parity-even spin-1 quartic tensor structures is given by
\begin{align*}
\mathbb{T}_1&=z_{14} z_{23}, \\
\mathbb{T}_2&= z_{13} z_{24},\\
\mathbb{T}_3& =z_{12} z_{34},\\
\mathbb{T}_4& =-zp_{32} zp_{42} z_{12}+zp_{31} zp_{43} z_{12}+zp_{13} zp_{21} z_{34}-zp_{14} zp_{24}
   z_{34},\\
\mathbb{T}_5&=zp_{32} zp_{42} z_{12}+zp_{32} zp_{43} z_{12}+zp_{14} zp_{21} z_{34}+zp_{14} zp_{24} z_{34},\\
\mathbb{T}_6& =zp_{31} zp_{42} z_{12}+zp_{13} zp_{24}
   z_{34},\\
\mathbb{T}_7& =zp_{24} zp_{42} z_{13}+zp_{13} zp_{31} z_{24},\\
\mathbb{T}_8& =zp_{14} zp_{31}
   z_{24}+zp_{13} zp_{32} z_{24}+zp_{21} zp_{43} z_{13}+zp_{14} zp_{32}
   z_{24}- zp_{24} zp_{42} z_{13},\\
\mathbb{T}_9& =zp_{21} zp_{31} z_{14}+zp_{13} zp_{43} z_{23}+zp_{21} zp_{32} z_{14}+zp_{14} zp_{43} z_{23},\\
\mathbb{T}_{10}& =zp_{24} zp_{31}
   z_{14}+zp_{13} zp_{42} z_{23},\\
\mathbb{T}_{11}& =zp_{21} zp_{42} z_{13}+zp_{24} zp_{42} z_{13}+zp_{21} zp_{43} z_{13}+zp_{24} zp_{43} z_{13}+zp_{14} zp_{32}
   z_{24},\\
\mathbb{T}_{12}& =zp_{21} zp_{32} z_{14}+zp_{24} zp_{32} z_{14}+zp_{14} zp_{42} z_{23}+zp_{14} zp_{43} z_{23},\\
\mathbb{T}_{13}& =zp_{13} zp_{24} zp_{31} zp_{42},\\
\mathbb{T}_{14}& =zp_{14} zp_{21} zp_{32} zp_{42}+zp_{14} zp_{24} zp_{32} zp_{42}+zp_{14} zp_{21} zp_{32} zp_{43}+zp_{14} zp_{24} zp_{32} zp_{43},\\
\mathbb{T}_{15}&=zp_{13} zp_{21} zp_{31} zp_{43}+zp_{14} zp_{21} zp_{31} zp_{43}+zp_{13} zp_{21} zp_{32} zp_{43}+zp_{14} zp_{21} zp_{32}
   zp_{43}, \\
\mathbb{T}_{16}&=zp_{14} zp_{24} zp_{31} zp_{42}+zp_{13} zp_{24} zp_{32}
   zp_{42}+ zp_{14} zp_{21} zp_{31} zp_{42}+zp_{13} zp_{24} zp_{32} zp_{43},\\
\mathbb{T}_{17}& =zp_{13}
   zp_{21} zp_{32} zp_{42}+zp_{14} zp_{24} zp_{31} zp_{43}+ zp_{13} zp_{21} zp_{31} zp_{42}+zp_{13} zp_{24} zp_{31} zp_{43}.
\end{align*}
A basis of the 10 $\mathbb{Z}_2^2$-invariant parity-odd spin-1 quartic tensor structures is given by
\begin{align*}
\tilde{\mathbb{T}}_{1}& = \varepsilon \left(z_1 z_2 z_3 z_4\right),\\
\tilde{\mathbb{T}}_{2}& =z_{34} \varepsilon \left(p_1 p_3 z_1 z_2\right)+z_{34} \varepsilon
   \left(p_2 p_3 z_1 z_2\right)+z_{12} \varepsilon \left(p_1 p_2 z_3 z_4\right),\\
\tilde{\mathbb{T}}_{3}& =z_{34} \varepsilon \left(p_1
   p_2 z_1 z_2\right)+z_{12} \varepsilon \left(p_1 p_3 z_3 z_4\right)+z_{12} \varepsilon \left(p_2 p_3 z_3
   z_4\right),\\
\tilde{\mathbb{T}}_{4}& =z_{24} \varepsilon \left(p_1 p_2 z_1 z_3\right)-z_{24} \varepsilon \left(p_2 p_3 z_1
   z_3\right)+z_{13} \varepsilon \left(p_1 p_3 z_2 z_4\right),\\
\tilde{\mathbb{T}}_{5}& =z_{14} \varepsilon \left(p_1 p_2 z_2 z_3\right)+z_{14}
   \varepsilon \left(p_1 p_3 z_2 z_3\right)-z_{23} \varepsilon \left(p_2 p_3 z_1 z_4\right),\\
\tilde{\mathbb{T}}_{6}& =z_{24} \varepsilon \left(p_1 p_3 z_1 z_3\right)+z_{13}
   \varepsilon \left(p_1 p_2 z_2 z_4\right)-z_{13} \varepsilon \left(p_2 p_3 z_2 z_4\right),\\
\tilde{\mathbb{T}}_{7}& =z_{23} \varepsilon \left(p_1 p_2 z_1
   z_4\right)+z_{23} \varepsilon \left(p_1 p_3 z_1 z_4\right)-z_{14} \varepsilon \left(p_2 p_3 z_2 z_3\right),\\
\tilde{\mathbb{T}}_{8}& =zp_{13} zp_{42} \varepsilon
   \left(p_1 p_2 z_2 z_3\right)+zp_{13} zp_{42} \varepsilon \left(p_1 p_3 z_2 z_3\right)-zp_{24} zp_{31} \varepsilon
   \left(p_2 p_3 z_1 z_4\right),\\
\tilde{\mathbb{T}}_{9}& = zp_{31} zp_{42} \varepsilon \left(p_1
   p_3 z_1 z_2\right)+zp_{31} zp_{42} \varepsilon \left(p_2 p_3 z_1 z_2\right)+zp_{13} zp_{24} \varepsilon \left(p_1 p_2
   z_3 z_4\right)
,\\
\tilde{\mathbb{T}}_{10}& =zp_{32} \left(zp_{42}+zp_{43}\right) \varepsilon \left(p_1 p_3 z_1 z_2\right)+zp_{32}
   \left(zp_{42}+zp_{43}\right) \varepsilon \left(p_2 p_3 z_1 z_2\right)+zp_{14} \left(zp_{21}+zp_{24}\right) \varepsilon
   \left(p_1 p_2 z_3 z_4\right).
\end{align*}

\section{Spin-1 Amplitude}
\label{app:solutions}
In this appendix we give the coefficients $f_I(s,t)$ and $\tilde{f}_{\tilde{I}}(s,t)$ of the spin-1 tensor structures, which are defined such that
\be
\mathcal{A}_{\rm contact} + \tilde{\mathcal{A}}_{\rm contact}  = \sum_{I=1}^{17} f_I(s,t) \mathbb{T}_I+ \sum_{\tilde{I}=1}^{10} \tilde{f}_{\tilde{I}}(s,t) \tilde{\mathbb{T}}_{\tilde{I}},
\ee
for the general amplitude that grows like $E^4$.
We set $m=1$ and use $s=2-2p_{12}$ and $t=2-2p_{13}$ as the Mandelstam variables for all incoming momenta. The parity-even coefficients are given by
\begin{align*}
f_1 =& 2 (4 \lambda_1-4 (2 + s - t) \lambda_2 + (s - t) (-\lambda_3 + 4 (4 + s - t) \lambda_4) + ((-4 + s) s + (-4 +t) t) \lambda_5 + 8 (2 \lambda_4 + \lambda_5)), \\ 
f_2 =&2 (4 \lambda_1 - 4 (-2 + t) \lambda_2 - (-4 + t) \lambda_3 + 4 (-2 + t)^2 \lambda_4 + 2 (4 + s^2) \lambda_5 + t (-4 - 2 s + t) \lambda_5), \\ 
f_3 =& 2 (4 \lambda_1  + 4 (-2 + s) \lambda_2 +s (\lambda_3 + 4 (-4 + s) \lambda_4) + (s (4 + s) - 2 (4 + s) t + 2 t^2) \lambda_5 + 8 (2 \lambda_4 + \lambda_5)), \\ 
f_4 =& -4 (2 \lambda_2 + 4 (-2 + s) \lambda_4 + s \lambda_5), \\ 
f_5 =& 2 (-4 \lambda_2 + \lambda_3 - 8 (-2 + s) \lambda_4 - 2 (6 + s - 2 t) \lambda_5), \\ 
f_6 =& f_8 =-2 (\lambda_3 + 4 (1 + s - t) \lambda_5), \\ 
f_7 =& 8 \lambda_2 - 2 (\lambda_3 + 8 (-2 + t) \lambda_4) + 4 (2 - 2 s + t) \lambda_5, \\ 
f_9 =& 2 (\lambda_3 + 4 (-3 + t) \lambda_5), \\ 
f_{10} =& f_{11}=-2 (\lambda_3 - 4 (-1 + s) \lambda_5), \\ 
f_{12} =& 4 (-2 \lambda_2 + 4 (2 + s - t) \lambda_4 + (s - t) \lambda_5), \\ 
f_{13} =& f_{14}=f_{15}= 8 (4 \lambda_4 + \lambda_5), \\ 
f_{16} =& f_{17} = 0.
\end{align*}
The parity-odd coefficients are given by\footnote{Naively we would expect $\tilde{f}_1=0$ so that the final amplitude is manifestly $S_4$ invariant when expressed in terms of tensor structures and Mandelstam variables. However, $S_4$ invariance is only required up to dimensionally dependent identities, which indeed holds in this case.}
\begin{align*}
\tilde{f}_1 =&16 (2 + s (-4 + t)) \tilde{\lambda}_2, \\ 
\tilde{f}_2 =& 32 (-3 + t) \tilde{\lambda}_2, \\ 
\tilde{f}_3 =& 16 (\tilde{\lambda}_1 + 2 (-3 + s) \tilde{\lambda}_2), \\ 
\tilde{f}_4 =& -32 (1 + s - t) \tilde{\lambda}_2, \\ \tilde{f}_5 =& 32 (-1 + s) \tilde{\lambda}_2, \\ 
\tilde{f}_6 =& 16 (\tilde{\lambda}_1 - 2 (-1 + t) \tilde{\lambda}_2), \\ 
\tilde{f}_7 =& -16 (\tilde{\lambda}_1 - 2 (3 + s - t) \tilde{\lambda}_2), \\ 
\tilde{f}_8 =&\tilde{f}_9 =\tilde{f}_{10} = -64 \tilde{\lambda}_2\,.
\end{align*}

\bibliographystyle{utphys}
\addcontentsline{toc}{section}{References}
\bibliography{amplitudes-arxiv-2}

\end{document}